\documentclass[aps, prd, amsmath, amssymb, twocolumn, 10pt, a4paper, superscriptaddress,nofootinbib,longbibliography]{revtex4-2}
\usepackage[percent]{overpic}
\usepackage{amsmath}
\usepackage{multirow}
\usepackage{booktabs}
\usepackage{array}
\usepackage{comment}
\usepackage{amssymb}
\usepackage{mathrsfs}
\usepackage{xcolor}
\usepackage{colortbl}
\usepackage{scalerel}
\usepackage{graphicx}
\usepackage{rotating}
\usepackage{float}
\usepackage{orcidlink}

\newcommand{\ie}{i.e.,~}

\usepackage{hyperref}
\hypersetup{colorlinks=true, linkcolor=blue, citecolor=cyan,
filecolor=blue, urlcolor=blue}

\begin{document}
\title{Viscosity effects on the shadow of a non-rotating black hole}

\author{Sayantani Lahiri\orcidlink{0000-0002-2187-5901}}
\email{sayantani.lahiri@gmail.com}
\affiliation{Institut für Theoretische Physik und Astrophysik, Universität Würzburg, Emil-Fischer-Strasse 31, 97074 Würzburg, Germany}
\thanks{From September 1, 2026: Chanakya University, Bengaluru 562164, Karnataka, India.}

\author{Sergio Gimeno-Soler\orcidlink{0000-0002-5352-4882}}
\email{sergio.gimeno@uv.es}
\affiliation{Departamento de Matem\'atica da Universidade de Aveiro and
Center for Research and Development in Mathematics and Applications (CIDMA),
Campus de Santiago, 3810-183 Aveiro, Portugal.}

\author{Alejandro Cruz-Osorio\orcidlink{0000-0002-3945-6342}}%
\email{aosorio@astro.unam.mx}
\affiliation{Instituto de Astronom\'{\i}a, Universidad Nacional Aut\'onoma de M\'exico, AP 70-264, Ciudad de M\'exico 04510, M\'exico}

\author{Jos\'e A. Font\orcidlink{0000-0001-6650-2634}}
\email{j.antonio.font@uv.es}
\affiliation{Departamento de Astronom\'{\i}a y Astrof\'{\i}sica, Universitat de Val\`encia, Avinguda Vicent 
Andrés Estellés 19, 46100, Burjassot (Val\`encia), Spain.}
\affiliation{Observatori Astron\`omic, Universitat de Val\`encia, C/ Catedr\'atico Jos\'e Beltr\'an 2, 46980, Paterna (Val\`encia), Spain.}

\author{Ziri Younsi\orcidlink{0000-0001-9283-1191}}
\email{z.younsi@ucl.ac.uk}
\affiliation{Mullard Space Science Laboratory, University College London, Holmbury St.~Mary, Dorking, Surrey, RH5 6NT, UK}

\begin{abstract}


We study the effect of shear viscosity in stationary magnetized accretion tori on synthetic images of non-rotating black hole shadows. Shear viscosity and spacetime-curvature contributions are introduced perturbatively in the tori through first and second-order transport coefficients within a second-order causal theory of non-ideal relativistic hydrodynamics. Synthetic black hole shadow images at 230\,GHz are obtained via general relativistic radiative transfer computations 
assuming thermal synchrotron emission and for a wide range of plasma magnetization parameters, viewing inclination angles, electron-temperature prescriptions, and viscosity parameters. A comparative pixel-by-pixel analysis using two normalized metrics shows that the largest image differences occur for strongly magnetized tori. While shear viscosity induces only minor changes in the overall shadow morphology, its combined effects with spacetime curvature are more evident in localized modifications of the synchrotron emission and pixel-wise flux distribution. These effects become increasingly pronounced at higher inclination angles, with the largest brightness differences between viscous and non-viscous configurations occurring for larger values of the electron-temperature parameter. Overall, our results indicate that shear viscosity and spacetime curvature leave only modest imprints on black hole shadow images produced by stationary thick disks, with differences remaining at the level of a few $\mu$Jy. Since our analysis is limited to stationary tori, the effects of shear viscosity might however be more significant in fully dynamical accretion systems.
\end{abstract}
    
\maketitle
\section{Introduction} \label{sec:intro}

Imaging the shadow of a supermassive black hole represents one of the 
most remarkable observational confirmations of general relativity in the 
strong-gravity regime. Millimeter-wavelength observations by the Event 
Horizon Telescope (EHT) Collaboration have revealed a robust silhouette 
characterized by a bright photon ring surrounding a central intensity 
depression, interpreted as the black hole shadow. This feature has been 
consistently observed over multiple years in both M87* and the Galactic 
center, Sagittarius\,A*, providing robust tests of gravitational physics in 
the vicinity of the event horizon \cite{EHT_M87_PaperI,EHT_M87_PaperV,
EHT_M87_2018,EHT_SgrA_PaperI,EHT_SgrA_PaperV_etal}. 

The black hole shadow has been studied for several decades, beginning 
with the pioneering work of \citet{1979AA....75..228L}. Subsequent studies 
explored the morphology of black hole images, such as the photon ring, 
shadow, and gravitational lensing effects, using analytical solutions in 
Schwarzschild \cite{Gralla:2019xty} and Kerr \cite{Abdujabbarov:2016hnw} spacetimes. These developments eventually led to the first 
connections with observational prospects \cite{Falcke:1999pj, Narayan:2013gca}.

The theoretical interpretation of the observed image has been extensively 
explored by considering black hole solutions beyond the Kerr spacetime and, more 
generally, beyond general relativity. Examples include Kerr-Newman-NUT 
black holes~\cite{Grenzebach:2014fha}, scenarios with extra 
dimensions~\cite{Vagnozzi:2019apd, Amarilla:2011fx}, and environments with plasma 
effects~\cite{Perlick:2015vta}. Alternative gravity frameworks have also been 
investigated, including Einstein-Gauss-Bonnet gravity~\cite{Galin:2021,Konoplya:2020bxa,
Cunha:2019dwb}, Einstein-dilaton-Gauss-Bonnet spacetimes
~\cite{Cunha:2016wzk}, and the influence of dark matter
~\cite{Konoplya:2019sns}. Additional studies include images of naked singularities~\cite{Galin:2019,Valentin:2025}, traversable wormholes~\cite{Valentin:2022}, photon ring properties 
in Yukawa-like black holes~\cite{Cruz2021}, boson stars~\cite{Olivares2020}, 
and dilaton black holes~\cite{Mizuno2018}, along with parametrized spacetimes 
such as the Rezzolla-Zhidenko and Johannsen-Psaltis metrics~\cite{Moriyama2025,
Wang2025,Younsi2016,Younsi2023,Chatterjee2025}.
Analyses have also been applied to a variety of charged black 
hole spacetimes~\cite{Kocherlakota2021,Chatterjee2023}.

Accurately modeling black hole shadow images reported by the EHT collaboration requires a proper description of the spacetime, which determines  
the size and morphology of the photon ring \cite{Johannsen2010,
EHT_M87_PaperVI,EHT_SgrA_PaperVI}. Moreover, the plasma model 
is equally crucial for understanding the detailed image morphology and 
the observed light curves. Astrophysical plasmas are typically described 
within the framework of magnetohydrodynamics and it is therefore essential 
to construct physically consistent initial configurations of magnetized plasma 
in magneto-centrifugal equilibrium around the black hole, as well as a 
realistic magnetic field structure.
Accretion disks surrounding astrophysical black holes provide an indirect 
probe of these compact objects through the process of accretion. In this 
process, angular momentum is transported outward while matter is drawn 
inward by the gravitational pull of the black hole, releasing a substantial 
amount of energy, part of which contributes to the disk luminosity 
\cite{2002apa..book.....F}, and converts gravitational and magnetic energy 
into particle acceleration and plasma heating.

Within the EHT collaboration, the most commonly used 
model for magnetized accretion disks is the Fishbone-Moncrief solution 
\cite{Fishbone76} (see also \cite{Kozlowski:1978,Abramowicz:1978,Uniyal2024}), 
which describes a hydrostatic equilibrium configuration with constant specific 
angular momentum; see also the solution by Font and Daigne~\cite{Font02a}. 
Extensions to non-constant angular momentum distributions were 
developed later by~\cite{Daigne04} (for a review, see \cite{Abramowicz:2011xu}). 
In these models, the magnetic field is typically introduced as an ad hoc 
perturbation, usually with a poloidal configuration, ensuring that the magnetic 
pressure remains much smaller than the gas pressure. 

A more self-consistent solution in magnetohydrostatic equilibrium with constant 
specific angular momentum was proposed by Komissarov \cite{Komissarov:2006}, 
commonly referred to as the Polish doughnut. Generalizations of this solution 
include equations of state with enthalpy dependence \cite{Gimeno_Soler_2017}, 
as well as the incorporation of magnetic polarization through magnetic susceptibility 
effects \cite{Pimentel2018a,Gimeno2024}.
General-relativistic magnetohydrodynamics 
(GRMHD) simulations based on Komissarov-type equilibria have been used to 
study jet launching and the development of magnetorotational instabilities 
\cite{Liska2018b,Pimentel2021}. A comparison between the Komissarov and 
Fishbone-Moncrief models can be found in \cite{Cruz2020}.

In our previous work \cite{Lahiri:2020sza}, we explored the modifications of Komissarov's solution by incorporating the effects of shear viscosity in magnetized thick accretion disks (or tori). In particular,
we constructed stationary solutions of viscous magnetized disks around a 
Schwarzschild black hole by employing the causal formulation of viscous relativistic hydrodynamics within the Eckart frame.
This approach was motivated by the well-known M\"uller-Israel-Stewart causal formulation \cite{Muller:1967zza, Israel:1976tn,ISRAEL1981204}. 
It was found that for highly magnetized plasmas, shear viscosity modifies the morphology of the torus by shifting the location of the cusp inward, thereby altering the equilibrium 
configuration and affecting the gas pressure distribution.

In this work, we compute images of black hole shadows of those stationary models, quantifying the effects of shear viscosity and including contributions that depend explicitly on the spacetime curvature. In order to assess the imprints of shear viscosity on the observed shadow we perform general relativistic radiative transfer (GRRT) calculations in the Schwarzschild spacetime, assuming thermal synchrotron 
emission. Our approach therefore provides a framework to probe non-ideal fluid effects in accretion flows.

This paper is organized as follows. Section \ref{sec:Tori} presents a summary of the procedure for building stationary equilibrium solutions of magnetized viscous tori around a Schwarzschild black hole. The basics of the GRRT calculations used to compute black hole shadow images from such viscous disks is discussed in Section \ref{sec:GRRT}. Our 
results are presented in Section \ref{sec:Shadows}, and the conclusions are summarized in Section \ref{sec:Summary}. Throughout the paper we use geometrized units $c=G=1$ and the metric is taken with the signature $\left(-,+,+,+\right)$.

    \section{Magnetized viscous tori} \label{sec:Tori}
    \subsection{Framework}
The solution for a magnetized viscous torus with a constant specific angular 
momentum distribution around a Schwarzschild black hole is obtained by 
solving the following conservation equations of relativistic ideal magnetohydrodynamics
\begin{equation}
  \nabla_\mu(\rho u^\mu)=0\,, \quad \nabla_\mu T_{\mathrm{ideal}}^{\mu \nu}=0\,, \quad \nabla_\mu {}^*F^{\mu \nu}=0\,, \label{eq:cons}
\end{equation}
where $\rho$ is the rest-mass density and $u^\mu$ is the fluid four-velocity. Here, 
we assume an axisymmetric flow, i.e., $u^{\mu} = (u^{t}, 0, 0, u^{\phi})$. In presence of non-zero shear viscosity, the 
energy-momentum tensor of the fluid in the Eckart frame is,
\begin{equation}
T^{\mu \nu}= (w+b^2)u^{\mu}u^{\nu}+\left(p+\frac{1}{2}b^2\right)g^{\mu\nu}-b^{\mu}b^{\nu}+\pi^{\mu \nu} \,,
\label{en-mom}
\end{equation}
where the enthalpy density is defined as $w = e + p$, with $p$ the fluid pressure 
and $e$ the total energy density. Using the gradient expansion scheme the shear viscosity tensor up to second-order gradients is given by \cite{Lahiri:2020sza,Lahiri_2020}
\begin{equation}
\noindent \pi^{\mu \nu}=-2\eta \sigma^{\mu \nu}+\tau_2 ^{<}D(2\eta \sigma^{\mu\nu})^{>} 
  +\kappa_2 u_{\alpha} u_{\beta} R^{\alpha <\mu \nu >\beta}\,. \label{viscosity}
\end{equation} 
Here $D\equiv u^{\alpha}\nabla_{\alpha}$ is the directional derivative and
$R^{\alpha\beta\gamma\delta}$ is the Riemann tensor, $\eta$ represents the 
shear viscosity coefficient whereas $\tau_2$ and $\kappa_2$ are the 
second-order transport coefficients. 
The angular brackets in Eq.~\eqref{viscosity} indicate traceless symmetric combinations, 
defined as
\begin{small}
\begin{eqnarray}
\sigma^{\mu \nu}&=&\triangle^{\mu \alpha}\triangle^{\nu \beta}\left(\displaystyle \frac{\nabla_{\alpha}u_{\beta}+\nabla_{\beta}u_{\alpha}}{2}\right)-\displaystyle \frac{1}{3} \triangle^{\mu \nu} \triangle^{\alpha \beta} \nabla_{\alpha}u_{\beta} \,, \nonumber \\
^{<}D\sigma^{\mu\nu}\,^{>}&=&\triangle^{\mu \alpha}\triangle^{\nu \beta}\left(\displaystyle \frac{D\sigma_{\alpha \beta}+D\sigma_{\beta \alpha}}{2}\right)\displaystyle -\frac{1}{3} \triangle^{\mu \nu} \triangle^{\alpha \beta}D\sigma_{\alpha \beta} \,, \nonumber \\ 
R^{\alpha <\mu \nu >\beta}&=&\left[\triangle^{\mu \rho} \triangle^{\nu \sigma}\left(\displaystyle \frac{  R^{\alpha}_{\;\rho \sigma \gamma}+ R^{\alpha}_{\; \sigma \rho \gamma}}{2}\right)-\frac{1}{3} \triangle^{\mu \nu} \triangle^{\rho \sigma}  R^{\alpha}_{\;\rho \sigma \gamma}\right]g^{\beta \gamma} \,, \nonumber
\end{eqnarray}
\end{small}
\noindent
where the projection tensor is defined as $\Delta^{\mu \nu} = g^{\mu \nu} + u^{\mu} u^{\nu}$. 
 
The Maxwell equations in relativistic ideal magnetohydrodynamics can be derived from Eq.~\eqref{eq:cons}, where the dual of the Faraday tensor~\cite{anile2005relativistic}, defined with respect to an observer with four-velocity $u^{\mu}$, is given by
 \begin{eqnarray}
  ^{*}F^{\mu \nu}= b^{\mu}u^{\nu}-b^{\nu}u^{\mu}\,,
  \end{eqnarray}
where $b^{\mu}$ is the magnetic field four-vector; here we assume a purely toroidal magnetic field, $b^{\mu} = (b^{t}, 0, 0, b^{\phi})$, which satisfies the orthogonality condition $u^{\alpha} b_{\alpha} = 0$. From this, we obtain the relation
   \begin{equation}
   b^2= (1-\Omega l_0)b^{\phi}b_{\phi}= 2p_{m } \label{pm}\,,
   \end{equation}
where $p_{m} \equiv b^{2}/2$ is the magnetic pressure, $l_0$ is the constant specific angular momentum of the torus, and $\Omega$ is the angular velocity of the fluid. The magnetic pressure can be compared with the gas pressure through the magnetization parameter, $\beta \equiv p/p_{m}$, which is independent of shear viscosity effects.
 
 The momentum conservation of energy-momentum tensor for the viscous fluid of the relativistic torus yields
 \begin{eqnarray}
(e+p)(u^{\rho}\nabla_{\rho}u_{\mu})+\triangle_{\mu}^ {\rho}\nabla_{\rho}\,p +\frac{\partial_{\mu}({\cal{L}}b^2)}{2{\cal{L}}}+ \nonumber \\g_{\mu \rho}\pi^{\rho \nu}(u^{\rho}\nabla_{\rho}u_{\nu}) 
+ \triangle_{\mu \gamma} \triangle_{\kappa \tau}\nabla^{\tau}\pi^{\gamma \kappa} &= &0 \,, 
\label{eq:full}
\end{eqnarray}
where ${\cal L} \equiv -g_{tt} g_{\phi \phi}$. This relation describes a viscous fluid in 
gravito-magneto-centrifugal equilibrium around a black hole.

We solve Eq.~\eqref{eq:full} by assuming that shear viscosity introduces linear perturbations 
in the fluid variables, namely the energy density, pressure, and magnetic field components, 
$w = (e, p, b^{t}, b^{\phi})$. Accordingly, we decompose $w = w_{0} + w_{(1)}$, where 
$w_{0}$ represents the background values and $w_{(1)} = (e_{(1)}, p_{(1)}, b^{t}_{(1)}, b^{\phi}_{(1)})$ 
denotes the first-order perturbations. 
We further assume $e = \rho (1 + \varepsilon) \approx \rho$, 
i.e., a cold plasma approximation. Consequently, Eq.~\eqref{eq:full} can be expressed in terms of the linear perturbation quantities as,
\begin{eqnarray}
(e_{(1)}+p_{(1)})a_{\mu}+ \frac{\partial_{\mu}\left[{\cal L} p_{m(1)} \right]}{{\cal L}}+\nonumber \\g_{\mu \rho}\pi^{\rho \nu}a_{\nu} 
+ \triangle_{\mu \gamma} \triangle_{\kappa \tau}\nabla^{\tau}\pi^{\gamma \kappa} &= &0 \label{eq:final}\,
\end{eqnarray}
where we adopt the equations of state $p = K e^{\gamma}$ and 
$p_{m} = K_m \mathcal{L}^{\gamma - 1} e^{\gamma}$, assuming the same polytropic index 
$\gamma$ for both pressure components \cite{Lahiri:2020sza, Komissarov:2006, Gimeno_Soler_2017}. 
Since shear viscosity is considered to have perturbative effects on solutions, following \citet{Lahiri:2020sza}, the transport coefficients $\eta$ and $\kappa_2$ are expressed in terms of the perturbation parameter $\lambda$. Thus, we have $\eta = \lambda m_1$ and $\kappa_2 = \lambda m_2$, where $m_1$ and $m_2$ are constants. Given that $\tau_2$ is the transport coefficient associated with the causality preserving term, without loss of generality, $\tau_2=1$ is set throughout this work.

Since the non-viscous disk quantities ($p_{(0)}$ and $e_{(0)}$) can be obtained using the standard procedure for inviscid disks, yielding algebraic expressions for these variables, we only need to solve a partial differential equation governing the first-order correction to the pressure, $p_{(1)}$.
\begin{equation}
\vec{\alpha}(r, \theta) \vec{\nabla}_{(r, \theta)} p_{(1)} - c(r, \theta) = 0\,.
\end{equation}
Further details regarding the construction of the models and results can be found in \cite{Lahiri:2020sza}.

\subsection{Magnetized viscous torus morphology}

The parameter space describing the morphology of a viscous, magnetized thick disk is fairly large. In order to keep it within a reasonable size we follow~\cite{Lahiri:2020sza} and we fix the value of the constant specific angular momentum of the disk to $l=3.8$ and the polytropic exponent to $\gamma = 5/3$. Moreover, we fix the value of the potential at the surface of the disk to $W_{\mathrm{s}} = -0.04$. This value yields a disk slightly overflowing its Roche lobe for our particular choice of $l$. In addition, we choose three different values for the magnetization parameter at the center of the disk $\beta_{\mathrm{c}}$, namely $100$, $10$ and $1$. Finally, we fix the two transport coefficients $m_1$ and $m_2$ by imposing that the absolute value of the first-order correction to the pressure is $80\%$ of its zeroth order value $|p_\mathrm{(1)}| = 0.8 p_{\mathrm{(0)}}$. Then, we apply this criterion to find a value of $m_1$ that fulfills that condition while keeping $m_2 = 0$. The same is done for $m_2$ when keeping $m_1 = 0$ and then again varying the two parameters. The complete set of parameters used in this work is reported in Table~\ref{table-parameters}.

\begin{table}[t]
\caption{Values of the transport coefficients $m_1$ and $m_2$ that are obtained for each value of $\beta_{\mathrm{c}}$ considered. We assume $W_{\mathrm{s}}=-0.04$ and $\Delta p_{\mathrm{cusp}} = (p_{(1)}/p_{(0)})|_{r_{\mathrm{cusp}}}= -0.8$.  Note that the cases with $m_1 = m_2 = 0$ are not included in the table.}   
\centering          
\begin{tabular}{c c c}
\hline\hline       
$m_1$ & $m_2$ & $\beta_{\mathrm{c}}$\\
\hline        
$6.46 \times 10^{-2} $ & $0$ & $10^{3}$  \\ 
$0$ & $1.48 \times 10^{-2}$ &   \\ 
$3.24 \times 10^{-2}$ & $7.41 \times 10^{-3}$ &   \\ 
\hline 
$7.07 \times 10^{-2} $ & $0$ & $10$  \\ 
$0$ & $1.61 \times 10^{-2}$ &   \\ 
$3.54 \times 10^{-2}$ & $8.09 \times 10^{-3}$ &   \\ 
\hline
$1.15 \times 10^{-1}$ & $0$ & $1$ \\ 
$0$ & $2.62 \times 10^{-2}$ &  \\ 
$5.78 \times 10^{-2}$ & $1.31 \times 10^{-2}$ &  \\ 
\hline
\end{tabular} 
\label{table-parameters}
\end{table}

\begin{figure}[t]
\includegraphics[width=1.0\linewidth]{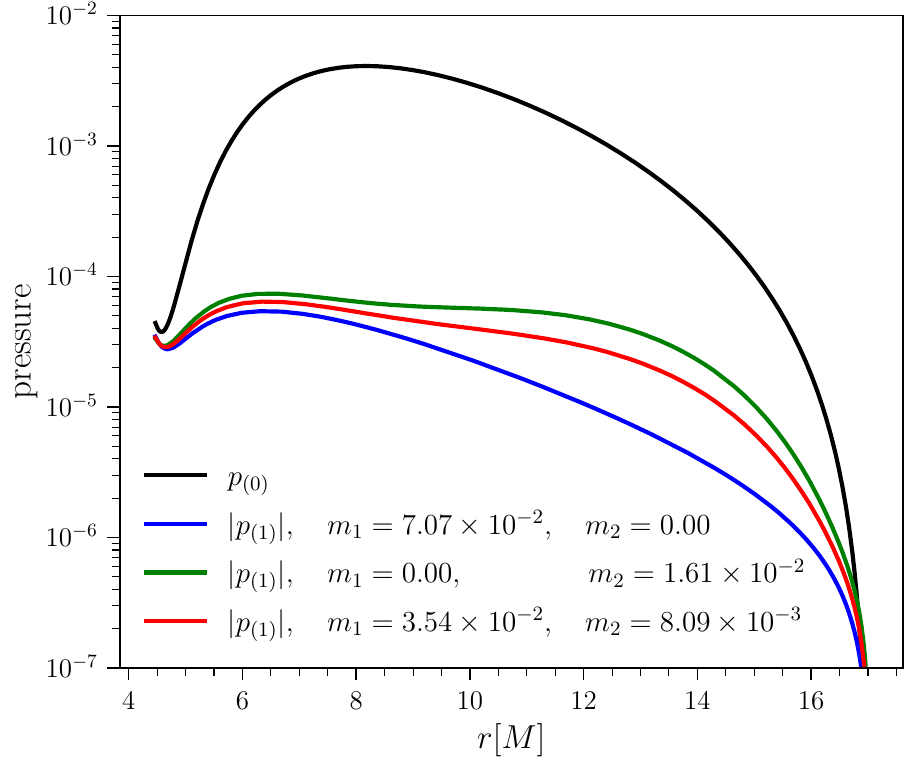}
\caption{Pressure, in logarithmic scale, as a function of the radial coordinate at the equatorial plane for the three models of Table~\ref{table-parameters} corresponding to $\beta_{\mathrm{c}} = 10$. The black line represents the zeroth order contribution to the pressure of the three models (i.e.~the pressure for a inviscid solution with $l=3.8$, $\beta_{\mathrm{c}} = 10$ and $W_{\mathrm{s}} = -0.04$). The blue line is the solution with $m_2 = 0$, the green line the solution with $m_1 = 0$, and the red line the solution in which both $m_1$ and $m_2$ are different from 0.}
\label{radial_plots}
\end{figure}

Figure~\ref{radial_plots} displays the radial distribution on the equatorial plane of $p_{(0)}$ and $|p_{(1)}|$ (in logarithmic scale) for the three sets of values $(m_1, m_2)$ corresponding to a value of the magnetization at the center of the disk of $\beta_{\mathrm{c}} = 10$. It can be seen that, even if the value of $|p_{(1)}|$ at the cusp is the same for all three models, $|p_{(1)}|$ behaves differently for each of them. The model in which $m_1 = 0$ and $m_2 \neq 0$ (green line) is the one with the highest deviation from the inviscid solution along the disk, followed by the model with $m_1$ and $m_2$ both non-vanishing (red line). Lastly, the model with $m_1 \neq 0$ and $m_2 = 0$ (blue line) yields the smaller deviations from the inviscid solution. 

This trend can also be seen in the upper half of each panel of Fig.~\ref{2D_plots}, which shows $|p_{(1)}|/|p_{(1),\mathrm{max}}|$ as a function of $(r, \theta)$. Here, it is apparent that the deviation from the inviscid solution is higher in general for the $m_1 = 0$ and $m_2 \neq 0$ model. This means that, almost everywhere, the total pressure $p_{\mathrm{t}} = p_{(0)} + p_{(1)}$ of this model is the lowest of the three choices that we consider for the viscosity parameters $(m_1, m_2)$ once $\beta_{\mathrm{c}}$ is fixed. The information contained in the lower half of the panels in Fig.~\ref{2D_plots} will be discussed in the following section. 

\section{General relativistic radiative transfer calculations}
\label{sec:GRRT}

\begin{figure*}
\includegraphics[width=1.00\linewidth]{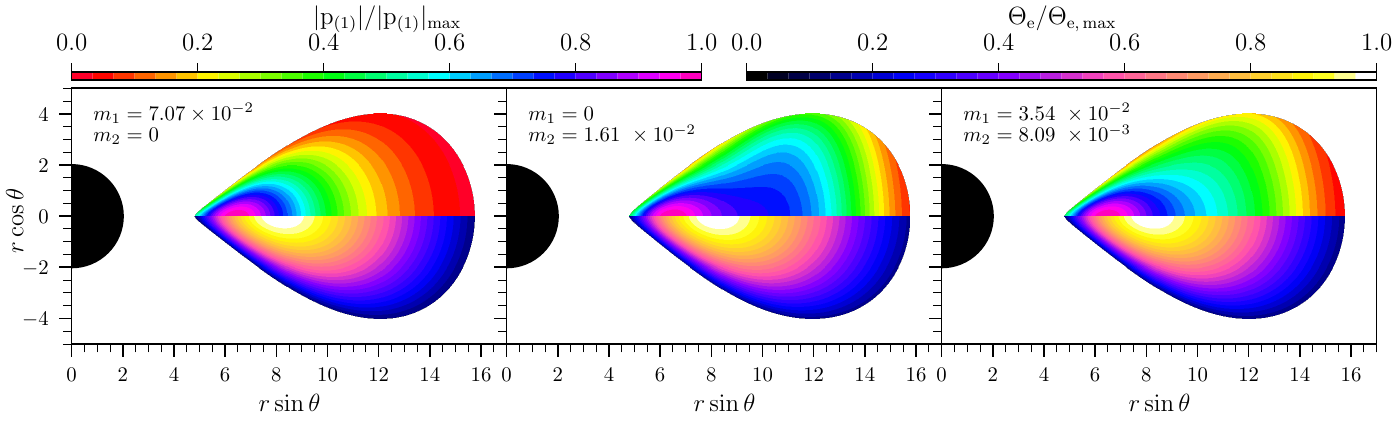}
\caption{Distribution of the absolute value of $p_{(1)}$ normalized to its maximum value, $(|p_{(1)}|/|p_{(1),\mathrm{max}}|$ (top half of each panel) and of the dimensionless electron temperature normalized to its maximum value $ \Theta_{\mathrm{e}}/\Theta_{\mathrm{e,max}}$ (bottom half of each panel). From left to right we show the models of Table~\ref{table-parameters} corresponding to  $\beta_{\mathrm{c}} = 10$. Note that the maximum values of the quantities $p_{(1),\mathrm{max}} = 7.384\times 10^{-5}$ and $\Theta_{{\mathrm{e,max}}}=4.067\times 10^{-3}$ can be used to recover the physical quantities. Additionally, the free parameters of the temperature ratio,  $R_{\mathrm{low}}$ and $R_{\mathrm{high}}$, are set to $R_{\mathrm{low}} = 1$ and $R_{\mathrm{high}} = 10$.}
\label{2D_plots}
\end{figure*}

To capture the radiative signatures across the electromagnetic spectrum of a 
viscous accretion disk in equilibrium around a Schwarzschild black hole, we 
solve the GRRT equations using the \texttt{BHOSS} code~\cite{Younsi:2019iee}. In covariant form, the radiative transfer equation can be written as,
\begin{equation}
    \frac{d\mathcal{I}}{d\tau_\nu}=-\mathcal{I} +\frac{\eta}{\chi},
\end{equation}
where $\mathcal{I}$ is the Lorentz-invariant specific intensity, related to the 
specific intensity $I_\nu$ by $\mathcal{I} = I_\nu / \nu^{3}$. The Lorentz-invariant emissivity $\eta$ and absorptivity $\chi$ are related to the emission and 
absorption coefficients, $j_\nu$ and $\alpha_\nu$, evaluated at frequency 
$\nu$, via $\eta = j_{0,\nu} / \nu^{2}$ and $\chi = \alpha_{0,\nu}\,\nu$. Here, 
the subscript “0” denotes quantities measured in the local rest frame of the 
plasma. Using the definition of the optical depth, $d\tau_\nu = \alpha_\nu\, ds$, 
the covariant radiative transfer equation can be rewritten as a system of two coupled differential equations,
\begin{eqnarray}
 \frac{d\tau_{\nu}}{d\lambda}= \gamma^{-1} \alpha_{0, \nu}\,, \quad  
 \frac{d{\mathcal{I} }}{d\lambda}= \gamma^{-1}  \left(\frac{j_{0,\nu}}{\nu^3}\right) \exp\left(-\tau_{\nu }\right), \label{eq:grrt}
\end{eqnarray}
with affine parameter $\lambda$ and energy shift between obervers and co-moving 
frame $\gamma^{-1}=\nu_0/\nu=-k_\alpha u^\alpha\rvert_\lambda/k_\beta u^{\,\beta}\rvert_\infty$ 
where $k_\alpha$ is the four-momentum. For more details on the radiative transfer 
scheme, see~\cite{Younsi2012}. The emission and absorption coefficients depend 
on the assumed emission process and electron distribution function, which we 
assume to be thermal, \ie a Maxwell-J\"uttner distribution given by 
\begin{equation}
\frac{dn_{\rm e}}{d\gamma_{\rm e}} = \frac{n_{\rm e}}{4 \pi \Theta_{\rm e}} \frac{\gamma_{\rm e} \sqrt{\gamma_{\rm e}^2 - 1}}{K_2\left(1/\Theta_{\rm e}\right)} \exp \left(- \frac{\gamma_{\rm e}}{\Theta_{\rm e}}\right) 
\label{eq:mjedf}, 
\end{equation}
where $\gamma_{e}$ is the electron Lorentz factor, $n_{\rm e}$ is the electron 
number density, $K_{2}$ is the Bessel function of second kind, and $\Theta_{e}$ 
is the dimensionless electron temperature (for 
more details see, e.g~\cite{Pandya2016}).

The viscous, magnetized accretion disk described in Section~\ref{sec:Tori} provides the proton temperature of the plasma. Therefore, 
the properties of the radiating electrons, particularly the electron temperature 
$T_e$, must be inferred from the plasma quantities. Here, we assume a 
plasma-$\beta$ dependent temperature ratio between protons and electrons, 
the so-called $R$–$\beta$ model \cite{Moscibrodzka2016,EHT_M87_PaperV,
Fromm2021b,Cruz2022,Cruz2026}, 
\begin{equation}
\Theta_{\rm e}= \frac{p m_{\rm p}/m_{\rm e} }{ \rho {\cal T}_{\rm ratio}},\,  \quad {\cal T}_{\rm ratio} \equiv \frac{T_{\rm p}}{T_{\rm e}}=\frac{R_{\rm low} + R_{\rm high} \beta^{2}}{1+\beta^{2}},
\label{eq:Te}
\end{equation}
where $m_{\rm p}$ and $m_{\rm e}$ are the proton and electron masses, respectively. 
The plasma-$\beta$ is obtained from our viscous torus solution. The temperature 
ratio depends on two additional free parameters, $R_{\rm low}$ and $R_{\rm high}$. 
In this prescription, $\beta \geq 1$ corresponds to weakly magnetized regions typically associated with the accretion disk, while $\beta < 1$ characterizes strongly magnetized regions (such as the polar jets emitted as the result of accretion processes).

The lower half of each panel of Fig.~\ref{2D_plots} displays the normalized dimensionless electron temperature $\Theta_{\mathrm{e}}/\Theta_{\mathrm{e, max}}$ for the three models of Table~\ref{table-parameters} with $\beta_{\mathrm{c}} = 10$. Following the definitions in~\citet{Lahiri:2020sza}, and noticing that the term of the dimensionless temperature that is affected by the presence of viscosity is just $p/\rho$, we rewrite Eq.~\eqref{eq:Te} in terms of the ratio between the first order correction to the pressure and the inviscid pressure, $\Delta p \equiv p_\mathrm{(1)}/p_{(0)}$,
\begin{equation}
    \Theta_{\mathrm{e}} \propto \frac{\gamma(1+\Delta p)}{\gamma + \Delta p}.
\end{equation}
This implies that models with more negative values of $\Delta p$, (i.e.~the one presented in the central column of Fig.~\ref{2D_plots}) have lower values of the dimensionless temperature, even if these differences are small. This is apparent  in Fig.~\ref{2D_plots}, where the differences between the various models are difficult to see with the naked eye.

We perform the GRRT calculations assuming M87 values for the black hole mass,  $6.5 \times 10^9\,M_\odot$, and distance, $16.8\,\mathrm{Mpc}$. The accretion rate is normalized to reproduce the observed total flux density of the M87* shadow at 230\,GHz, namely $\simeq 1.0\,\mathrm{Jy}$ \citep{Akiyama2015,Doeleman2012}. We adopt a field of view of $30\,M$ and an image resolution of $512 \times 512$ pixels. Furthermore, we explore a set of models by varying key physical parameters, including the central plasma magnetization parameter $\beta_{\mathrm{c}}$, the electron temperature parameter $R_{\rm high}$, the viewing inclination angle $\iota$, and, in particular, the effects of shear viscosity through the $m_1$ and $m_2$ coefficients.

\section{Shadows of viscous tori}
\label{sec:Shadows}

To assess the impact of shear viscosity on shadow images, we compare the shadow structures produced by viscous and non-viscous disks across different values of the system parameters. To generate the images, the same values of $l_0$ and $\gamma$ are used irrespective of the viscosity, ensuring a consistent comparison. While keeping $R_{\mathrm{low}}=1$ fixed, the magnetization parameter $\beta_{\mathrm{c}}$ is varied from moderately magnetized ($\beta_{\mathrm{c}}=1$) to weakly magnetized ($\beta_{\mathrm{c}}=10^3$) disk configurations. For each value of $\beta_{\mathrm{c}}$, the viscosity parameters $m_1$ and $m_2$ are chosen such that the perturbative treatment remains valid (see Table~\ref{table-parameters}) and for the corresponding value of $W_s$. The shadow images are then generated for the full set of allowed values of $m_1$, $m_2$, and $\beta_{\mathrm{c}}$, while systematically varying the inclination angle $\iota$ and the parameter $R_{\mathrm{high}}$, which characterizes the electron--proton temperature ratio.


Simultaneously varying all model parameters when generating shadow images provides limited insight, particularly when attempting to disentangle the effects of shear viscosity. Therefore, among the three variables $\beta_{\mathrm{c}}$, $R_{\mathrm{high}}$, and the inclination angle $\iota$, we fix two and vary the third. The resulting shadow images are then compared between a viscous disk ($m_i \neq 0, i=1,2$) and its non-viscous counterpart ($m_i=0$) in order to isolate the individual effects of $\eta$ and $\kappa_2$ through $m_1$ and $m_2$. To further facilitate the comparison of the emitted spectral flux density, we compute the pixel-wise intensity difference, $\Delta S= S_{\mathrm{non-viscous}} - S_{\mathrm{viscous}}$, for each model. In all case studies, the observing frequency is fixed at 230 GHz, and the maximum intensity $I_{\mathrm{max}}$ is reported for each configuration.

\subsection{Effect of the inclination view angle}

\begin{figure}[t]
    \includegraphics[width=0.5\textwidth]{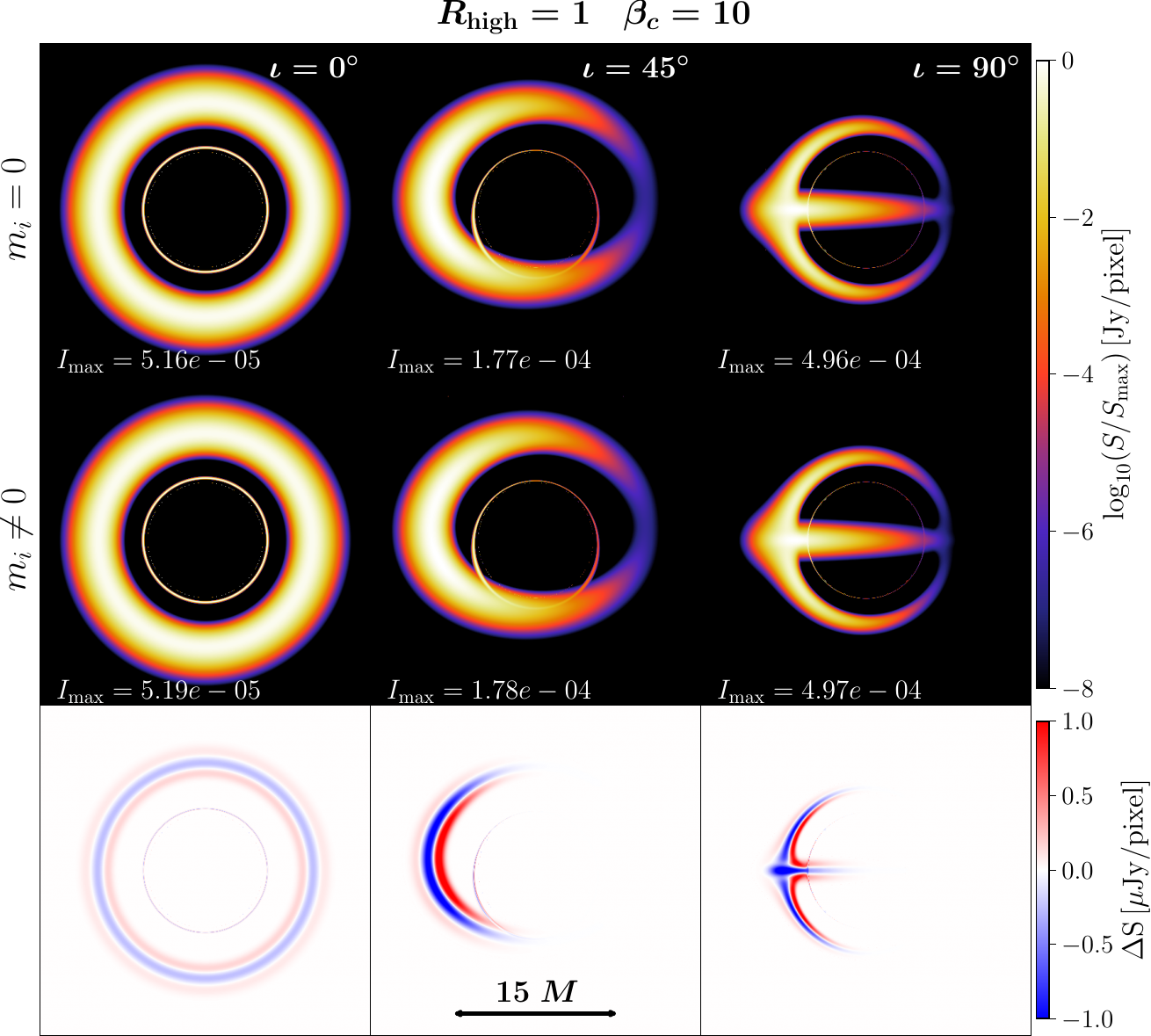}
    \caption{Black hole shadows at $230\, {\rm GHz}$ for three inclination view angle $\iota=0^{\circ}, 45^{\circ}, 90^{\circ}$ (from left to right). The first row shows  Komissarov's solution without viscous effects. The second row shows the viscosity effects when $m_{1} \neq 0$ and  $m_{2} \neq 0$. The third row shows the differences between the viscous and non-viscous solutions.}
\label{rp1}
\end{figure}

 We consider three inclination angles, namely $\iota= 0^{\circ}, 45^{\circ}, 90^{\circ}$ in equal steps covering the full range of viewing angles. Fig.~\ref{rp1} shows a comparison of the shadow images when the black hole is illuminated by a magnetized non-viscous torus (first row) and a magnetized viscous torus (second row), keeping $R_{\mathrm{high}}=1$ and $\beta_{\mathrm{c}}=10$ (we refer to Table~\ref{table-parameters} for the exact values of $m_1$ and $m_2$). The comparison between the first and second horizontal panels shows that the overall morphology of the images remains unchanged under variations of the inclination angle including the face-on view ($\iota=0°$) and the edge-on view ($\iota=90°$). The effects of the shear viscosity and the spacetime curvature can be identified from the third row which highlights them in terms of difference in pixel-wise spectral flux density $\Delta S$ from the viscous and non-viscous solutions. When viewed face-on, one obtains an uniform distribution of difference in spectral density per pixel throughout the disk. In contrast,  for higher inclination angles, the distribution of $\Delta S$ develops increasingly prominent red and blue regions, particularly on the approaching side towards the black hole.
 This behavior indicates that shear viscosity and curvature-induced modifications of the synchrotron emission become more visible in near edge-on viewing geometries. This implies that with increasing viewing angles, the synchrotron emission corresponding to the viscous source becomes more pronounced in some regions of the disk in comparison to the non-viscous source thereby giving rise to systematic distribution of suppression and enhancement regions in $\Delta S$. 
 We also find that $I_{\mathrm{max}}$ remains nearly unchanged between the magnetized viscous and non-viscous configurations.

\subsection{Effect of the electron temperature} 

\begin{figure}[t]
    \includegraphics[width=0.5\textwidth]{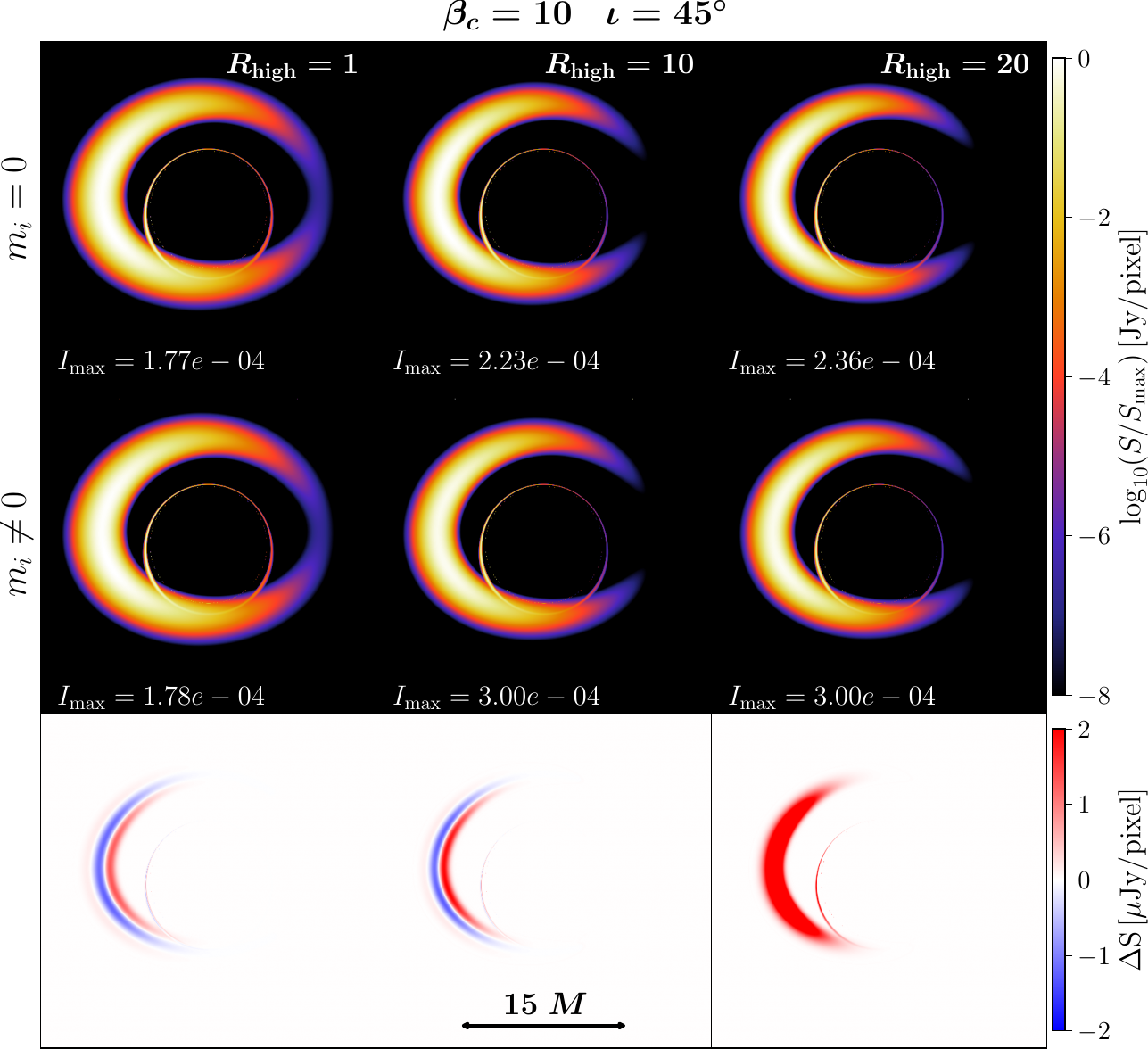}
    \caption{Same as Fig.~\ref{rp1} but for three values of the electron temperature parameter $R_{\rm high}=1$, 10, and 20 (from left to right).}
\label{rp2}
\end{figure}

Next, we fix the values of the magnetization parameter and inclination angle. We consider a moderately magnetized torus with $\beta_{\mathrm{c}}=10$ and  set the viewing  angle at $\iota=45^{\circ}$  while systematically increasing the electron temperature parameter as $R_{\mathrm{high}}=1$, 10, and 20. Fig.~\ref{rp2} illustrates the effects of varying the electron temperature parameter. As before, the first horizontal panel shows shadows due to a non-viscous torus whereas the second panel refers to  a viscous magnetized torus with $m_{i}\neq 0, i=1,2$. The third row displays the pixel-wise spectral flux density difference between the two cases. We note that by increasing $R_{\mathrm{high}}$, the overall spectral flux density (normalized by $S_{\mathrm{max}}$) diminishes gradually. However, this does not significantly modify the overall morphology of the shadow images.
On the other hand, the spatial distribution of the synchrotron emission undergoes noticeable changes, particularly in the surrounding of the shadow boundary. This is shown by $\Delta S$, where localized red and blue regions progressively develop near the brightest emitting regions as the parameter $R_{\mathrm{high}}$ increases. In the presence of viscosity, variations of $R_{\mathrm{high}}$ not only modify the electron temperature through the adopted $R-\beta$ prescription (note that with $\beta_{\mathrm{c}}=10$, $R_{\mathrm{low}}=1$, $T_e \propto R_{\mathrm{high}}^{-1} $), but also alter how viscosity-induced changes in the plasma structure are imprinted onto the synchrotron emission. 
For the moderately magnetized configuration ($\beta_{\mathrm{c}}=10$), larger values of $R_{\mathrm{high}}$ correspond to lower electron temperatures in the disk-dominated regions. At the same time, the viscous correction to the pressure remains negative which implies that shear viscosity locally reduces the total plasma pressure relative to the non-viscous configuration.

The combined effect of low-temperature electrons and shear viscosity and curvature-induced pressure suppression leads to increasingly pronounced spatial differences in the emitted radiation. In particular, the red regions correspond to locations where the viscous magnetized torus emits less strongly than the non-viscous configuration. Conversely, the blue regions indicate localized enhancement of emission due to shear viscosity and curvature contributions to the emitting plasma.
The increasing prominence of these structures with larger $R_{\mathrm{high}}$ therefore suggests that shear viscosity and curvature leave stronger imprints on the emission when the electrons within the disk becomes colder.
Furthermore, the maximum intensity $I_{\mathrm{max}}$ is enhanced with increasing $R_{\mathrm{high}}$ when non-viscous disks are considered. 


\subsection{Effect of the plasma magnetization} 

\begin{figure}[t]
    \includegraphics[width=0.5\textwidth]{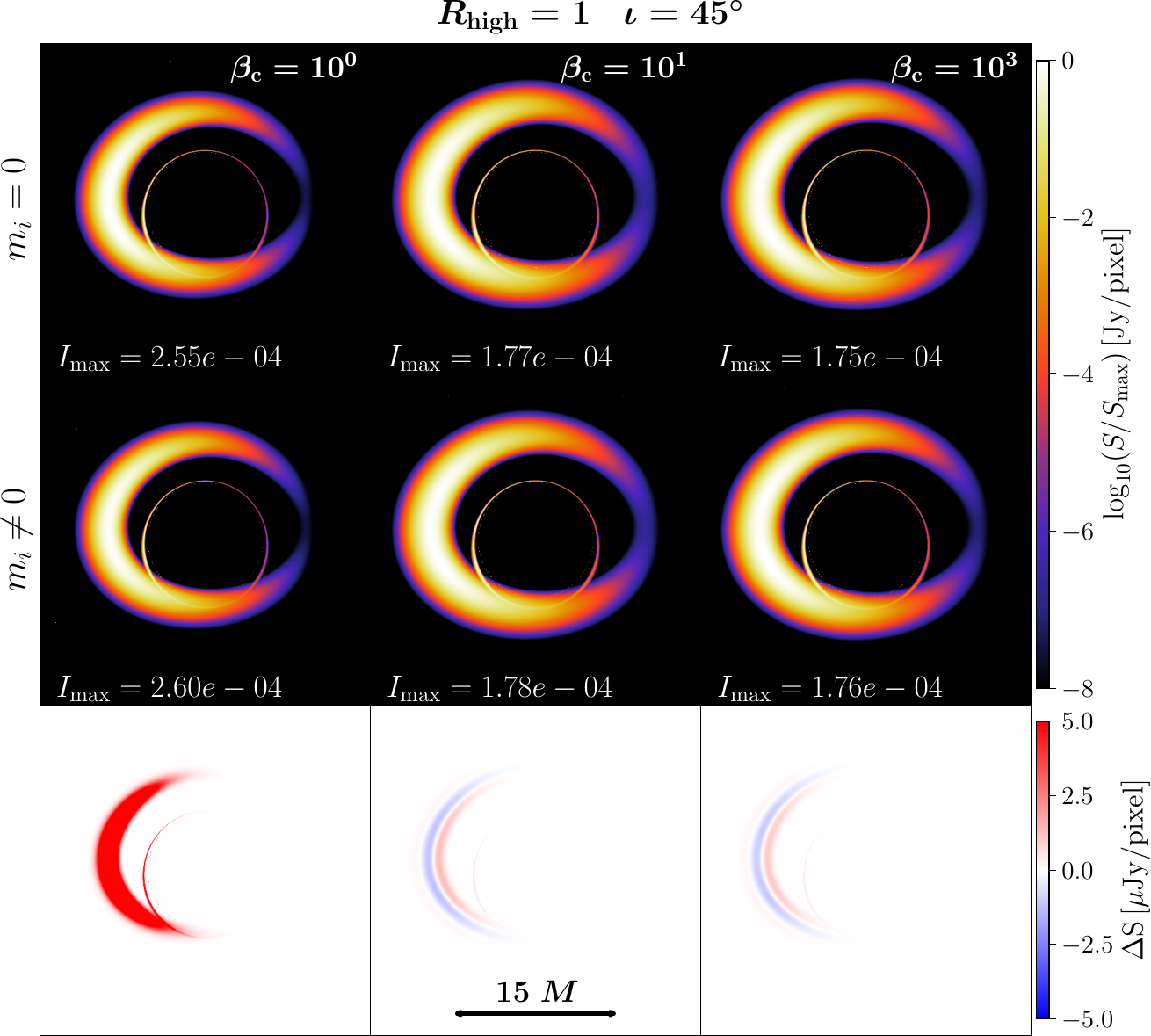}\\
    \caption{Same as Fig.~\ref{rp1} but for three values of the plasma magnetization parameter $\beta_{\mathrm{c}}=1$, 10, and $10^{3}$ (from left to right).}
\label{rp3}
\end{figure}

Fig.~\ref{rp3} illustrates the shadow images when varying the plasma magnetization parameter $\beta_{\mathrm{c}}$ and the shear viscosity with non-vanishing $m_i$. In this case we fix $R_{\mathrm{high}}=1$ and the inclination angle to $\iota=45°$. The shadow images in Fig.~\ref{rp3} span from moderately magnetized disks ($\beta_{\mathrm{c}}=1$) to weakly magnetized counterparts ($\beta_{\mathrm{c}}=10^3$).  The images in the first and second rows in Fig.~\ref{rp3} demonstrate that, with varying magnitude of $\beta_{\mathrm{c}}$, the morphological changes of the shadows between inviscid and viscous models are noticeably small.

The pixel-wise spectral flux difference is displayed in the third row.  The choice of parameters leads to ${\cal T}_{\mathrm{ratio}} =1$ everywhere within the torus. This implies that varying $\beta_{\mathrm{c}}$ does not change the electron temperature and any changes in $\Delta S$ cannot be attributed to temperature effects. Therefore, nontrivial imprints on the synchrotron emission are due to shear viscosity in this case. As the emission strongly depends on the disk magnetization, the differences observed in $\Delta S$ purely reflect shear-viscosity and magnetization-induced effects  rather than changes associated with the electron temperature-dependent effects.

We find that for the highest magnetized disk  ($\beta_{\mathrm{c}}=1$) $\Delta S$ is enhanced as shown by the extended red regions over a large fraction of the image. This indicates that the synchrotron emission from the viscous torus is systematically suppressed relative to the non-viscous configuration. However, as $\beta_{\mathrm{c}}$ increases, corresponding to progressively weakly magnetized torus, $\Delta S$ becomes increasingly localized and structured, with both red and blue regions appearing across different parts of the shadow. This suggests that viscosity produces spatially non-uniform modifications of the synchrotron emission, leading to local suppression as well as enhancement of the pixel-wise flux density depending on the plasma conditions within the torus. In particular, the positive $\Delta S$ regions become weakest for the weakly magnetized configuration, implying that the suppression of emission due to shear viscosity becomes less pronounced in such disks. A similar trend is also observed for the blue regions, indicating that viscosity-induced enhancement of the emission likewise diminishes with decreasing magnetization. Overall, the flux difference $\Delta S$ indicates that the contrast between the viscous and non-viscous flux distributions becomes progressively less pronounced for weakly magnetized configurations. Finally, it is to be noted that for the particular set of disk parameters considered in this section, viscosity and curvature effects contribute to a slight enhancement of $I_{\mathrm{max}}$, noticeable for  $\beta_{\mathrm{c}}=1$.

 
\subsection{Effect of the shear viscosity} 

\begin{figure}[t]
    \includegraphics[width=0.5\textwidth]{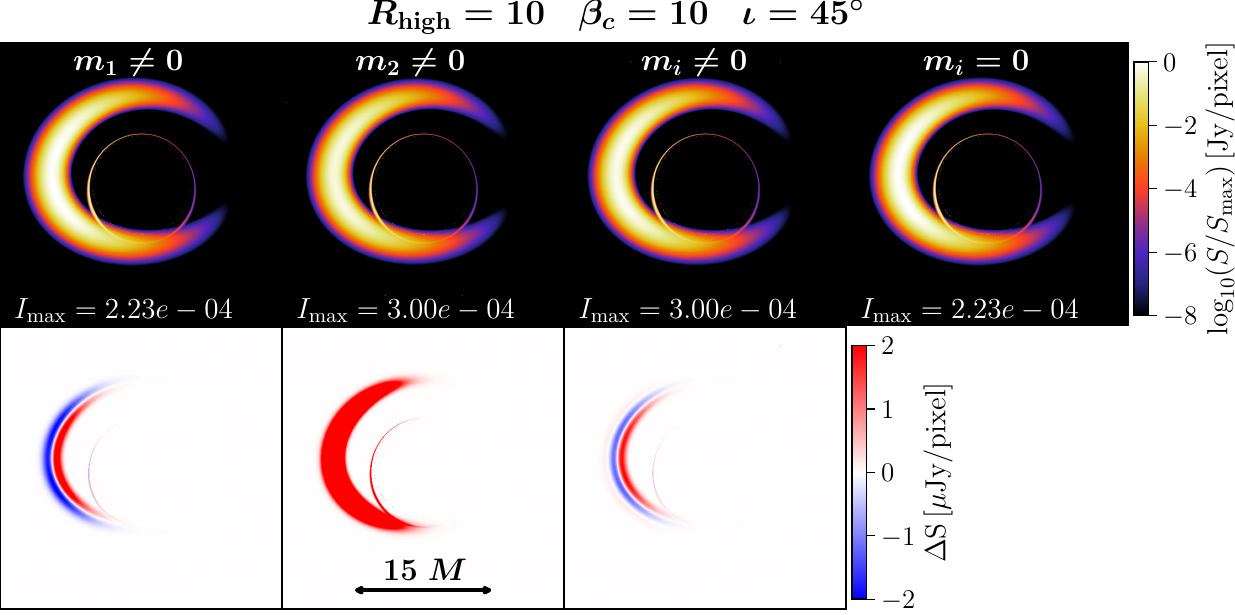}\\
    \caption{Same as Fig.~\ref{rp1} but for different values of the transport coefficients (from left to right $m_{1} \neq 0$, $m_{2} \neq 0$, $m_{i} \neq 0$ , and $m_{i} =0$). }
\label{rp4}
\end{figure}

 We now analyse the inidividual impact of the two transport coefficients $m_1$ and $m_2$ on the shadow images, keeping fixed the viewing inclination angle, $\iota=45^{\circ}$, the disk plasma magnetization, $\beta_{\mathrm{c}}=10$,  and the temperature, $R_{\mathrm{high}}=10$. The results are shown in Fig.~\ref{rp4}, using the values for $m_1$ and $m_2$ from Table~\ref{table-parameters}.
 The last column of the first row corresponds to the shadow image from a non-viscous magnetized torus whereas the first three colmuns correspond to the images obtained by the presence of a viscous magnetized disk. We note that for the model parameters considered, together with the $R-\beta$ parametrization, the electron temperature is relatively lower than the proton temperature throughout the disk. This observation is important in interpreting the flux emission in the presence of shear viscosity.    
 
While overall morphological changes in the shadow images are less recognizable between non-viscous and viscous cases, as noticed in the top row of Fig.~\ref{rp4}, the pixel-wise flux difference $\Delta S$ reveals distinct pattens depending on which viscous contribution is switched on. Those involve the causality-preserving transport coefficient (represented by $m_1$), the transport coefficient representing spacetime curvature (i.e.~$m_2$), and their interplay ($m_i \neq 0$). The impact in each case differs because shear viscosity modifies the plasma structure in distinct ways, depending on the physical properties of the specific transport coefficient under consideration. In the following, we discuss the effects of varying $m_1$ and $m_2$ on the shadow images.

\textit{First column of Fig.~\ref{rp4} - $m_1\neq 0$}:
In this case $\Delta S$ exhibits both red and blue regions distributed around the crescent structure. The red and blue patches are relatively balanced, indicating that the causality-preserving shear viscosity term produces both local suppression and enhancement of the emission depending on the emitting region.

\textit{Second column of Fig.~\ref{rp4} - $m_2\neq 0$}:
This case is dominated by an extended red region of $\Delta S$, implying that emission from the viscous magnetized torus is less strong than the non-viscous one over larger portions of the image. It therefore suggests that the corresponding spacetime curvature contribution systematically modifies the plasma in a way that suppresses synchrotron emission more globally than in the $m_1\neq 0$ case.
At the same time, the slight increase in $I_{\mathrm{max}}$ indicates that the remaining emission becomes more concentrated near the brightest regions.

 \textit{Third column of Fig.~\ref{rp4} - $m_i\neq 0$}:
When the contributions of both transport coefficients are simultaneously present, the resulting $\Delta S$ pattern becomes localized. The causality preserving shear viscosity and the spacetime curvature terms both accounting for the total shear viscosity tensor partially compete with each other, producing local cancellation in some regions and reinforcement in others. Consequently, the observable flux redistribution becomes less pronounced in comparison to the $m_1\neq 0$ case.

In summary we find that the low electron temperature enhances the sensitivity of  emission to overall shear viscosity, while the different sub-terms modify the plasma structure in qualitatively different ways, thereby producing distinct $\Delta S$ patterns in the three cases.

\begin{figure*}[t]
	\includegraphics[scale=0.65]{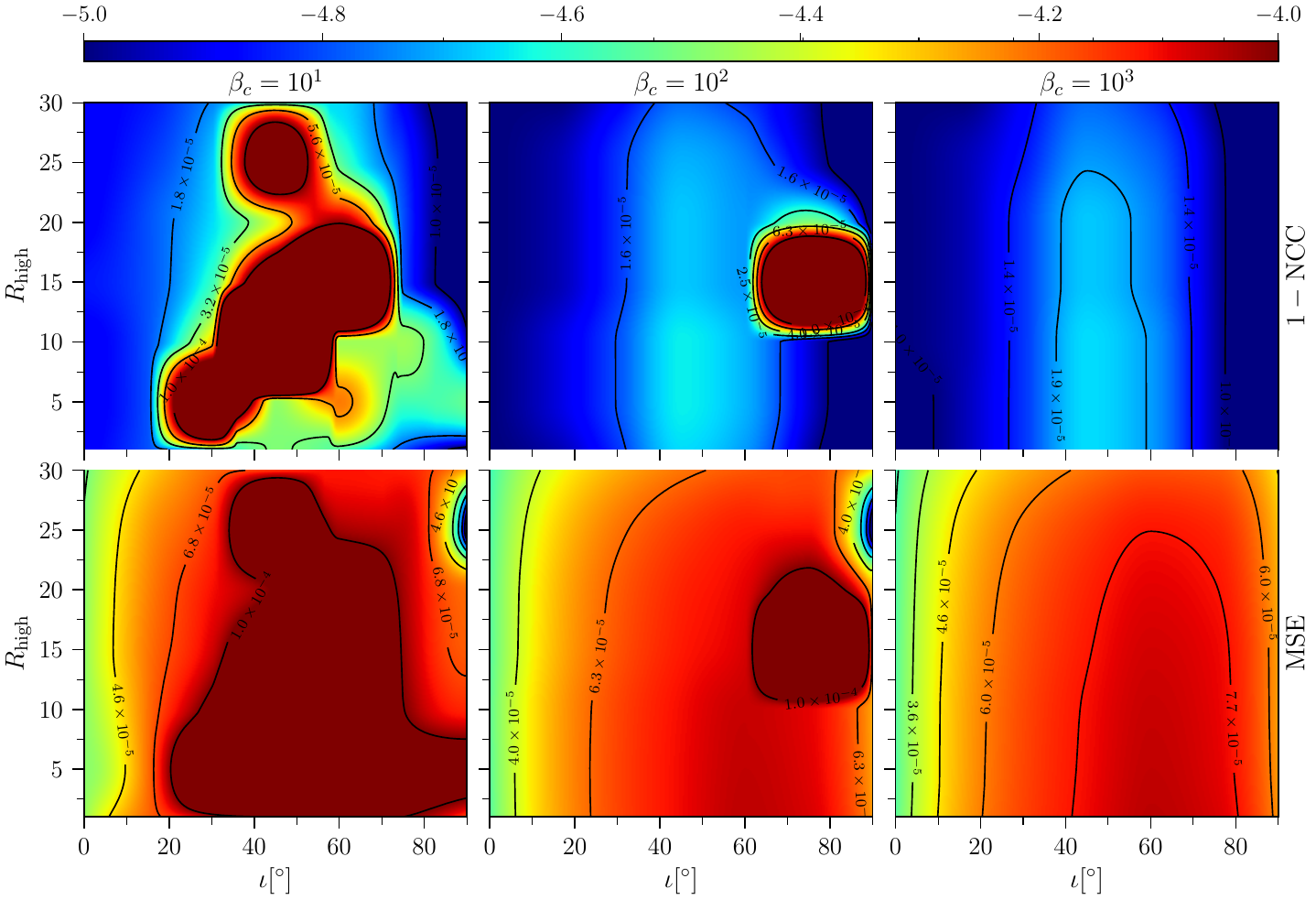}\\
	\caption{Pixel-to-pixel image comparison of the inclination viewing angle $\iota$ vs $R_{\mathrm{high}}$ using two image comparison metrics, 1-NCC (top panel) and MSE (bottom panel). In all cases $m_1 \neq0, m_2 \neq0$. The different columns correspond to different values of the plasma magnetization parameter, $\beta_{\mathrm{c}}=10$, $10^2$, and $10^3$ (from left to right).}
	\label{com_plot1}
\end{figure*}

\subsection{Comparison of images with different metrics }

To further investigate the signatures of the transport coefficients in the shadow images, we extract and analyze quantitative image features employing a method that  quantifies the pixel-to pixel differences between two types of shadows in the image plane following the procedure presented in \citet{Roeder2023}.  
We consider two standard image comparison metrics, namely normalized cross-correlation ($\mathrm{NCC}$) and the mean squared error ($\mathrm{MSE}$). It is known that the normalized correlation metric gives rise to unity for two exactly identical images and decreases as the images become less similar \cite{1284395, Mizuno:2018lxz}.
For convenience we report $1-\mathrm{NCC}$ as a measure of dissimilarity. Hence, a value of 0 indicates identical images. On the other hand, in our case, the mean squared error quantifies the average pixel-wise intensity difference between the viscous and non-viscous images \cite{995823}.
This normalized definition yields a dimensionless error measure: $\mathrm{MSE}=0$ if the images are identical, with higher values indicating larger total intensity discrepancies (with the normalization ensuring that the metric is independent of the absolute brightness scale). Hence the image comparison technique enables to assess how viscosity affects the image brightness distribution. As we show below, in our comparisons the $\mathrm{MSE}$ is on the order of $10^{-4}$ or less which indicate that shear viscosity induces only very slight overall changes in image brightness. 

 For each set of fixed physical parameters, we  compare images produced with a magnetized viscous torus to the corresponding image with an inviscid torus. The results of the analysis are presented in 
Fig.~\ref{com_plot1}. The two metrics, $1-\mathrm{NCC}$ and $\mathrm{MSE}$, are displayed across the full range of inclination angles and $R_{\rm high}$ for three sets of magnetization parameter $\beta_{\mathrm{c}}$. The values of the latter are purposefully chosen to cover different intensities of magnetization of the torus. The color maps in Fig.~\ref{com_plot1}  are represented using a logarithmic scale in order to resolve variations spanning approximately one order of magnitude.

 For $\beta_{\mathrm{c}} = 10$ (left column), the distribution of $1-\mathrm{NCC}$ exhibits localized maxima at two distinct regions of the parameter space. The strongest deviations occur at moderate-to-high inclination angles and low-to-moderate values of $R_{\mathrm{high}}$. A second enhancement appears at large $R_{\mathrm{high}}$ ($R_{\mathrm{high}} \gtrsim 20$). The contour structure reveals that the image similarity varies by nearly an order of magnitude across the parameter space, indicating a strong sensitivity of the shadow morphology to the electron-temperature prescription. The corresponding $\mathrm{MSE}$ map displays a broadly similar trend, with the largest errors concentrated in the same region of moderate and high inclinations. However, unlike $1-\mathrm{NCC}$, the $\mathrm{MSE}$ distribution is smoother and occupies a larger fraction of the parameter space. 

Changing the plasma magnetization parameter to $\beta_{\mathrm{c}} = 10^2$ (middle column) substantially alters the structure of the maps. In the $1-\mathrm{NCC}$ panel, the broad region of enhanced deviations collapses into a small region centered approximately at $\iota \sim 60^\circ$--$75^\circ$ and $R_{\mathrm{high}} \sim 10$--$15$. Outside this narrow region, the pixel-to-pixel difference remains close to its minimum value throughout the parameter space. 
 The $\mathrm{MSE}$ panel likewise develops a localized enhancement in approximately the same region, however, unlike the $1-\mathrm{NCC}$ map, it also retains a smooth background variation across the parameter space. In particular, away from the localized peak, the metric varies predominantly with the viewing inclination, whereas its dependence on $R_{\mathrm{high}}$ is comparatively weak. Thus, unlike $\mathrm{1-NCC}$ metric comparisons, which primarily captures localized structural differences, the $\mathrm{MSE}$ metric remains sensitive to the broader redistribution of pixel intensities, causing the pixel-wise intensity differences between the viscous and non-viscous images to become increasingly pronounced at larger viewing inclinations.
It may be noted that both metrics exhibit their strongest enhancements at approximately the same location in parameter space. Also, $\beta_{\mathrm{c}} = 10^2$ marks an intermediate regime in which the influence of the plasma prescription is still observable but is confined to specific combinations of the inclination angle and $R_{\mathrm{high}}$.

For $\beta_c = 10^3$ (right column), the localized enhancement present in the $\beta_{\mathrm{c}}=10^2$ maps disappears entirely, and both diagnostics exhibit smooth large-scale gradients across the parameter space. In the $1-\mathrm{NCC}$ panel, the diagnostic increases gradually from low inclinations toward intermediate inclinations, reaching its largest values around $\iota\sim45^\circ$--$60^\circ$, before decreasing again at larger viewing angles. At fixed inclination, variations with $R_{\mathrm{high}}$ remain comparatively small, indicating that the inclination angle is the primary driver of the residual image differences.
A similar behavior is observed in the $\mathrm{MSE}$ map. The error increases systematically from low to intermediate inclinations and then decreases toward larger inclinations, producing broad nearly vertical contour bands. The comparatively weak curvature of the contours demonstrates that variations in $R_{\mathrm{high}}$ induce only minor modifications to the diagnostics relative to those produced by changing the viewing angle.

Overall, our analysis indicates that the observable imprint of viscosity and spacetime curvature is strongest in highly magnetized accretion flows. For the most magnetized configuration ($\beta_{\mathrm{c}}=10$), both $1-\mathrm{NCC}$ and $\mathrm{MSE}$ exhibit extended regions of enhanced deviations, demonstrating a sensitivity of the shadow images to the plasma prescription. As the magnetization decreases ($\beta_{\mathrm{c}}=10^2$), these broad regions reduces. For the weakest magnetization considered ($\beta_{\mathrm{c}}=10^3$), the localized enhancement disappears entirely and the diagnostics vary smoothly across the parameter space. The results therefore suggest that strong magnetization amplifies the observable impact of viscosity and curvature on shadow images, whereas weakly magnetized flows tend to suppress these signatures and leave only modest, large-scale variations dominated by viewing angles.

\section{Summary and discussion}\label{sec:Summary}

We have studied possible observational signatures of shear viscosity in magnetized accretion tori in  images of black hole shadows. Following~\cite{Lahiri:2020sza}, the impact of shear viscosity on the images has been studied by introducing viscosity contributions perturbatively through first and second-order transport coefficients associated with shear viscosity and spacetime-curvature in a second-order causal theory of non-ideal relativistic hydrodynamics. This approach allows to determine curvature effects on equilibrium solutions of stationary magnetized viscous tori through those transport coefficients. We have investigated this procedure in the particular physical system of a Schwarzschild black hole surrounded by a magnetized thick disk with a constant angular momentum distribution. 
 
Using general relativistic radiative transfer (GRRT) calculations and assuming thermal synchrotron emission, we have generated synthetic shadow images for a range of plasma magnetization parameters, viewing inclination angles, electron-temperature prescriptions, and viscosity parameters. In the adopted GRRT framework, synchrotron emissivity depends on the electron temperature, as inferred from the plasma properties through the $R-\beta$ prescription. Therefore, any shear viscosity-induced changes in the total pressure distribution in the disk can potentially alter the electron temperature and leave observable imprints on the emitted radiation. Our analysis has revealed that for a fixed plasma magnetization $\beta_c$, the curvature-dependent transport coefficient $m_2$ produces the largest deviations from the inviscid solution (see Fig.~\ref{radial_plots}), whereas coefficient $m_1$ yields comparatively weaker modifications. As a result, the total pressure gets reduced relative to the inviscid configuration. This reduction leads to systematically lower electron temperatures, although the temperature variations remain modest. These results demonstrate that the curvature-related viscous contribution has the strongest impact on the thermodynamic properties of the disk plasma thereby providing favourable conditions for observable signatures through  GRRT calculations. 

Conducting such computations we have observed that shear viscosity produces only minor changes in the overall morphology of the black hole shadow. However, the cumulative effects of viscosity and spacetime curvature are more clearly visible through localized modifications of the synchrotron emission and pixel-wise flux distributions.
Those effects become increasingly more apparent for larger inclination angles, where differences between viscous and inviscid models are amplified. The effects of plasma magnetization are more pronounced, with the 
highly magnetized configurations exhibiting the strongest differences between viscous and non-viscous images. Furthermore, the adopted electron-temperature prescription also plays an important role, with larger values of $R_{\mathrm{high}}$ producing stronger brightness differences between viscous and non-viscous configurations. In addition, we have observed that curvature-induced viscosity effects tend to produce stronger and more extended suppression of synchrotron emission, whereas the causality-preserving contribution leads to a more balanced redistribution of the emitted flux.

These observations have been corroborated from a comparative study of pixel-by-pixel image differences using two normalized metrics, cross-correlation and mean square error. Both metric show that the largest image differences occur for strongly magnetized accretion flows. As the magnetization decreases, these differences become increasingly localized and eventually reduce to smooth, weak variations mostly controlled by the viewing angle geometry. Overall, this study has demonstrated that shear viscosity and spacetime curvature leave modest imprints on images of black hole shadows sourced by stationary thick disks, with the strongest observational signatures arising in fairly high magnetized systems.  In all cases, however, the differences between viscous and inviscid models remain at the level of a few $\mu$Jy.

This work represents a first step toward understanding the potential impact of dissipative effects on black hole shadow images within a simple and controlled framework in a non-rotating spacetime. In this study, viscosity has been incorporated as a perturbative effect in stationary models of equilibrium magnetized relativistic tori. Building on this work, several directions for further investigation remain open. These, for example, include the computation of shadow images in fully dynamical accretion systems, the inclusion of a non-zero radial velocity component, and the consideration of non-uniform angular momentum distributions within the tori, which are particularly relevant for angular momentum transport mediated by shear viscosity. An even more realistic extension would be to study these effects in the spacetime of a rotating black hole, both for stationary configurations and through the subsequent time evolution of such initial data. We leave these investigations to future work.

\section*{Acknowledgements}
SL acknowledges the hospitality of Lehrstuhl für Astronomie, Julius Maximilian University of Würzburg where part of this work has been completed. 
ACO is supported by DGAPA-UNAM grant IN110522, the Ciencia Básica y de Frontera 2023–2024 program of SECIHTI México projects CBF2023-2024-1102, 257435 and 1147615. JAF acknowledges support from the Spanish Agencia Estatal de Investigación (grant PID2024-159689NB-C21) funded by MICIU/AEI/10.13039/501100011033 and by FEDER / EU, from the Generalitat Valenciana (Prometeo Excellence Programme grant CIPROM/2022/49), and from the European Horizon Europe staff exchange (SE) programme HORIZON-MSCA-2021-SE-01 (Grant No.-NewFunFiCO-101086251). Simulations were performed at Atocatl LAMOD-UNAM cluster, and ``Going Merry" workstation at IA-UNAM. 
\bibliographystyle{apsrev4-1}
\bibliography{references}

@ARTICLE{995823,
  author={Zhou Wang and Bovik, A.C.},
  journal={IEEE Signal Processing Letters}, 
  title={A universal image quality index}, 
  year={2002},
  volume={9},
  number={3},
  pages={81-84},
  doi={10.1109/97.995823}}

@ARTICLE{1284395,
  author={Zhou Wang and Bovik, A.C. and Sheikh, H.R. and Simoncelli, E.P.},
  journal={IEEE Transactions on Image Processing}, 
  title={Image quality assessment: from error visibility to structural similarity}, 
  year={2004},
  volume={13},
  number={4},
  pages={600-612},
  doi={10.1109/TIP.2003.819861}}

@article{Akiyama2015,
        Adsurl = {http://adsabs.harvard.edu/abs/2015ApJ...807..150A},
        Archiveprefix = {arXiv},
        Author = {{Akiyama}, K. and {Lu}, R.-S. and {Fish}, V.~L. and {Doeleman}, S.~S. and {Broderick}, A.~E. and {Dexter}, J. and {Hada}, K. and {Kino}, M. and {Nagai}, H. and {Honma}, M. and {John\
son}, M.~D. and {Algaba}, J.~C. and {Asada}, K. and {Brinkerink}, C. and {Blundell}, R. and et al.},
        Doi = {10.1088/0004-637X/807/2/150},
        Eid = {150},
        Eprint = {1505.03545},
        Journal = {Astrophys. J.},
        Month = jul,
        Pages = {150},
        Primaryclass = {astro-ph.HE},
        Title = {{230 GHz VLBI Observations of M87: Event-horizon-scale Structure during an Enhanced Very-high-energy {$\gamma$}-Ray State in 2012}},
        Volume = 807,
        Year = 2015}

@ARTICLE{Galin:2019,
       author = {{Gyulchev}, Galin and {Nedkova}, Petya and {Vetsov}, Tsvetan and {Yazadjiev}, Stoytcho},
        title = "{Image of the Janis-Newman-Winicour naked singularity with a thin accretion disk}",
      journal = {\prd},
         year = 2019,
        month = jul,
       volume = {100},
       number = {2},
          eid = {024055},
        pages = {024055},
          doi = {10.1103/PhysRevD.100.024055},
archivePrefix = {arXiv},
       eprint = {1905.05273},
 primaryClass = {gr-qc},
       adsurl = {https://ui.adsabs.harvard.edu/abs/2019PhRvD.100b4055G}
}

@ARTICLE{Galin:2021,
       author = {{Gyulchev}, Galin and {Nedkova}, Petya and {Vetsov}, Tsvetan and {Yazadjiev}, Stoytcho},
        title = "{Image of the thin accretion disk around compact objects in the Einstein-Gauss-Bonnet gravity}",
      journal = {European Physical Journal C},
         year = 2021,
        month = oct,
       volume = {81},
       number = {10},
          eid = {885},
        pages = {885},
          doi = {10.1140/epjc/s10052-021-09624-5},
archivePrefix = {arXiv},
       eprint = {2106.14697},
 primaryClass = {gr-qc},
       adsurl = {https://ui.adsabs.harvard.edu/abs/2021EPJC...81..885G}
}

@ARTICLE{Valentin:2022,
       author = {{Deliyski}, Valentin and {Gyulchev}, Galin and {Nedkova}, Petya and {Yazadjiev}, Stoytcho},
        title = "{Polarized image of equatorial emission in horizonless spacetimes: Traversable wormholes}",
      journal = {\prd},
         year = 2022,
        month = nov,
       volume = {106},
       number = {10},
          eid = {104024},
        pages = {104024},
          doi = {10.1103/PhysRevD.106.104024},
archivePrefix = {arXiv},
       eprint = {2206.09455},
 primaryClass = {gr-qc},
       adsurl = {https://ui.adsabs.harvard.edu/abs/2022PhRvD.106j4024D}
}

@ARTICLE{Valentin:2025,
       author = {{Deliyski}, Valentin and {Gyulchev}, Galin and {Nedkova}, Petya and {Yazadjiev}, Stoytcho},
        title = "{Observing naked singularities with the present and next-generation Event Horizon Telescope}",
      journal = {\prd},
         year = 2025,
        month = mar,
       volume = {111},
       number = {6},
          eid = {064068},
        pages = {064068},
          doi = {10.1103/PhysRevD.111.064068},
archivePrefix = {arXiv},
       eprint = {2401.14092},
 primaryClass = {gr-qc},
       adsurl = {https://ui.adsabs.harvard.edu/abs/2025PhRvD.111f4068D}
}

@ARTICLE{Uniyal2024,
       author = {{Uniyal}, Akhil and {Dihingia}, Indu K. and {Mizuno}, Yosuke},
        title = "{A Revisited Equilibrium Solution of the Fishbone and Moncrief Torus for Extended General Relativistic Magnetohydrodynamic Simulations}",
      journal = {Astrophys. J.},
         year = 2024,
        month = aug,
       volume = {970},
       number = {2},
          eid = {172},
        pages = {10},
          doi = {10.3847/1538-4357/ad5b5b},
archivePrefix = {arXiv},
       eprint = {2406.16309},
 primaryClass = {astro-ph.HE},
       adsurl = {https://ui.adsabs.harvard.edu/abs/2024ApJ...970..172U}
}

@article{Doeleman2012,
        Adsurl = {http://adsabs.harvard.edu/abs/2012Sci...338..355D},
        Archiveprefix = {arXiv},
        Author = {{Doeleman}, S.~S. and {Fish}, V.~L. and {Schenck}, D.~E. and {Beaudoin}, C. and {Blundell}, R. and {Bower}, G.~C. and {Broderick}, A.~E. and {Chamberlin}, R. and {Freund}, R. and {F\
riberg}, P. and {Gurwell}, M.~A. and {Ho}, P.~T.~P. and {Honma}, M. and {Inoue}, M. and {Krichbaum}, T.~P. and et al.},
	Doi = {10.1126/science.1224768},
        Eprint = {1210.6132},
        Journal = {Science},
        Month = oct,
        Pages = {355-},
        Primaryclass = {astro-ph.HE},
        Title = {{Jet-Launching Structure Resolved Near the Supermassive Black Hole in M87}},
        Volume = 338,
        Year = 2012}

@ARTICLE{Daigne04,
       author = {{Daigne}, Fr{\'e}d{\'e}ric and {Font}, Jos{\'e} A.},
        title = "{The runaway instability of thick discs around black holes - II. Non-constant angular momentum discs}",
      journal = {Mon. Not. R. Astron. Soc.},
         year = 2004,
        month = apr,
       volume = {349},
       number = {3},
          eid = {841-868},
        pages = {841-868},
          doi = {10.1111/j.1365-2966.2004.07547.x},
archivePrefix = {arXiv},
       eprint = {astro-ph/0311618},
 primaryClass = {astro-ph},
       adsurl = {https://ui.adsabs.harvard.edu/abs/2004MNRAS.349..841D}
}

@ARTICLE{Moriyama2025,
       author = {{Moriyama}, Kotaro and {Cruz-Osorio}, Alejandro and {Mizuno}, Yosuke and {Dihingia}, Indu K. and {Uniyal}, Akhil},
        title = "{Black hole accretion and radiation variability in general relativistic magnetohydrodynamic simulations with Rezzolla{\textendash}Zhidenko spacetime}",
      journal = {Astron. Astrophys.},
         year = 2025,
        month = feb,
       volume = {694},
          eid = {A135},
        pages = {A135},
          doi = {10.1051/0004-6361/202452679},
archivePrefix = {arXiv},
       eprint = {2501.08720},
 primaryClass = {astro-ph.HE},
       adsurl = {https://ui.adsabs.harvard.edu/abs/2025A&A...694A.135M}
}

@article{Mizuno2018,
       author = {{Mizuno}, Yosuke and {Younsi}, Ziri and {Fromm}, Christian M. and
         {Porth}, Oliver and {De Laurentis}, Mariafelicia and
         {Olivares}, Hector and {Falcke}, Heino and {Kramer}, Michael and
         {Rezzolla}, Luciano},
        title = "{The current ability to test theories of gravity with black hole shadows}",
      journal = {Nature Astronomy},
         year = 2018,
        month = apr,
       volume = {2},
        pages = {585-590},
          doi = {10.1038/s41550-018-0449-5},
archivePrefix = {arXiv},
       eprint = {1804.05812},
 primaryClass = {astro-ph.GA},
       adsurl = {https://ui.adsabs.harvard.edu/abs/2018NatAs...2..585M}
}

@article{Mizuno:2018lxz,
    author = "Mizuno, Yosuke and Younsi, Ziri and Fromm, Christian M. and Porth, Oliver and De Laurentis, Mariafelicia and Olivares, Hector and Falcke, Heino and Kramer, Michael and Rezzolla, Luciano",
    title = "{The Current Ability to Test Theories of Gravity with Black Hole Shadows}",
    eprint = "1804.05812",
    archivePrefix = "arXiv",
    primaryClass = "astro-ph.GA",
    doi = "10.1038/s41550-018-0449-5",
    journal = "Nature Astron.",
    volume = "2",
    number = "7",
    pages = "585--590",
    year = "2018"
}

@ARTICLE{Cruz2022,
       author = {{Cruz-Osorio}, Alejandro and {Fromm}, Christian M. and {Mizuno}, Yosuke and {Nathanail}, Antonios and {Younsi}, Ziri and {Porth}, Oliver and {Davelaar}, Jordy and {Falcke}, Heino and {Kramer}, Michael and {Rezzolla}, Luciano},
       title = "{State-of-the-art energetic and morphological modelling of the launching site of the M87 jet}",
       journal = {Nature Astronomy},
       year = 2022,
       month = jan,
       volume = {6},
       pages = {103-108},
       doi = {10.1038/s41550-021-01506-w},
       archivePrefix = {arXiv},
       eprint = {2111.02517},
       primaryClass = {astro-ph.HE},
       adsurl = {https://ui.adsabs.harvard.edu/abs/2022NatAs...6..103C}
}

@ARTICLE{Cruz2026,
       author = {{Cruz-Osorio}, Alejandro and {Meringolo}, Claudio and {Fromm}, Christian M. and {Mizuno}, Yosuke and {Servidio}, Sergio and {Nathanail}, Antonios and {Younsi}, Ziri and {Rezzolla}, Luciano},
        title = "{Supermassive Black Hole Imaging with a Self-consistent Electron-temperature Prescription}",
      journal = {\apj},
         year = 2026,
        month = apr,
       volume = {1001},
       number = {2},
          eid = {227},
        pages = {227},
          doi = {10.3847/1538-4357/ae4b3d},
archivePrefix = {arXiv},
       eprint = {2512.14835},
 primaryClass = {astro-ph.HE},
       adsurl = {https://ui.adsabs.harvard.edu/abs/2026ApJ..1001..227C}
}

@ARTICLE{Roeder2023,
       author = {{R{\"o}der}, Jan and {Cruz-Osorio}, Alejandro and {Fromm}, Christian M. and {Mizuno}, Yosuke and {Younsi}, Ziri and {Rezzolla}, Luciano},
        title = "{Probing the spacetime and accretion model for the Galactic Center: Comparison of Kerr and dilaton black hole shadows}",
      journal = {\aap},
         year = 2023,
        month = mar,
       volume = {671},
          eid = {A143},
        pages = {A143},
          doi = {10.1051/0004-6361/202244866},
archivePrefix = {arXiv},
       eprint = {2301.09549},
 primaryClass = {astro-ph.HE},
       adsurl = {https://ui.adsabs.harvard.edu/abs/2023A&A...671A.143R}
}

@article{Younsi2012,
	Adsurl = {http://adsabs.harvard.edu/abs/2012A%26A...545A..13Y},
	Archiveprefix = {arXiv},
	Author = {{Younsi}, Z. and {Wu}, K. and {Fuerst}, S.~V.},
	Doi = {10.1051/0004-6361/201219599},
	Eid = {A13},
	Eprint = {1207.4234},
	Journal = {Astron. Astrophys.},
	Month = sep,
	Pages = {A13},
	Primaryclass = {astro-ph.HE},
	Title = {{General relativistic radiative transfer: formulation and emission from structured tori around black holes}},
	Volume = 545,
	Year = 2012}

@article{Younsi:2019iee,
    author = "Younsi, Ziri and Porth, Oliver and Mizuno, Yosuke and Fromm, Christian M. and Olivares, Hector",
    title = "{Modelling the polarised emission from black holes on event horizon-scales}",
    eprint = "1907.09196",
    archivePrefix = "arXiv",
    primaryClass = "astro-ph.HE",
    doi = "10.1017/S1743921318007263",
    journal = "IAU Symp.",
    volume = "342",
    pages = "9--12",
    year = "2020"
}

@ARTICLE{Pandya2016,
       author = {{Pandya}, Alex and {Zhang}, Zhaowei and {Chandra}, Mani and {Gammie}, Charles F.},
        title = "{Polarized Synchrotron Emissivities and Absorptivities for Relativistic Thermal, Power-law, and Kappa Distribution Functions}",
      journal = {Astrophys. J.},
         year = 2016,
        month = may,
       volume = {822},
       number = {1},
          eid = {34},
        pages = {34},
          doi = {10.3847/0004-637X/822/1/34},
archivePrefix = {arXiv},
       eprint = {1602.08749},
 primaryClass = {astro-ph.HE},
       adsurl = {https://ui.adsabs.harvard.edu/abs/2016ApJ...822...34P} }

@ARTICLE{Pimentel2018a,
        Author = {{Pimentel}, O.~M. and {Lora-Clavijo}, F.~D. and {Gonzalez}, G.~A.},
        Title = "{Analytic solution of a magnetized tori with magnetic polarization around Kerr black holes}",
        Journal = {Astron. Astrophys.},
        ArchivePrefix = "arXiv",
        Eprint = {1808.07400},
        PrimaryClass = "astro-ph.HE",
        Year = 2018,
        Month = nov,
        Volume = 619,
        Eid = {A57},
        Pages = {A57},
        Doi = {10.1051/0004-6361/201833736},
        Adsurl = {http://adsabs.harvard.edu/abs/2018A%26A...619A..57P}}

@ARTICLE{Cruz2020,
        Author = {{Cruz-Osorio}, Alejandro and {Gimeno-Soler}, Sergio and {Font}, Jos{\'e} A.},
        Title = "{Non-linear evolutions of magnetized thick discs around black holes: dependence on the initial data}",
        Journal = {Mon. Not. R. Astron. Soc.},
        Year = "2020",
        Month = mar,
        Volume = {492},
        Number = {4},
        Pages = {5730-5742},
        Doi = {10.1093/mnras/staa216},
        ArchivePrefix = {arXiv},
        Eprint = {2001.09669},
        PrimaryClass = {gr-qc},
        Adsurl = {https://ui.adsabs.harvard.edu/abs/2020MNRAS.492.5730C}}

@article{Cruz2021,
       author = "Cruz-Osorio, Alejandro and Gimeno-Soler, Sergio and Font, Jos\'e A. and De Laurentis, Mariafelicia and Mendoza, Sergio",
       title = "{Magnetized discs and photon rings around Yukawa-like black holes}",
       eprint = "2102.10150",
       archivePrefix = "arXiv",
       primaryClass = "astro-ph.HE",
       doi = "10.1103/PhysRevD.103.124009",
       journal = "Phys. Rev. D",
       volume = "103",
       number = "12",
       pages = "124009",
       year = "2021"
}

@article{Cunha:2016wzk,
    author = "Cunha, Pedro V. P. and Herdeiro, Carlos A. R. and Kleihaus, Burkhard and Kunz, Jutta and Radu, Eugen",
    title = "{Shadows of Einstein\textendash{}dilaton\textendash{}Gauss\textendash{}Bonnet black holes}",
    eprint = "1701.00079",
    archivePrefix = "arXiv",
    primaryClass = "gr-qc",
    doi = "10.1016/j.physletb.2017.03.020",
    journal = "Phys. Lett. B",
    volume = "768",
    pages = "373--379",
    year = "2017"
}

@article{Konoplya:2019sns,
    author = "Konoplya, R. A.",
    title = "{Shadow of a black hole surrounded by dark matter}",
    eprint = "1905.00064",
    archivePrefix = "arXiv",
    primaryClass = "gr-qc",
    doi = "10.1016/j.physletb.2019.05.043",
    journal = "Phys. Lett. B",
    volume = "795",
    pages = "1--6",
    year = "2019"
}

@article{Kocherlakota2021,
	author = {Kocherlakota, Prashant and others},
	collaboration = {Event Horizon Telescope},
	title = "{Constraints on black-hole charges with the 2017 EHT observations of M87*}",
	eprint = {2105.09343},
	archivePrefix = {arXiv},
	primaryClass = {gr-qc},
	doi = {10.1103/PhysRevD.103.104047},
	journal = {Phys. Rev. D},
	volume = {103},
	number = {10},
	pages = {104047},
	year = 2021}

@article{Cunha:2019dwb,
    author = "Cunha, Pedro V. P. and Herdeiro, Carlos A. R. and Radu, Eugen",
    title = "{Spontaneously Scalarized Kerr Black Holes in Extended Scalar-Tensor\textendash{}Gauss-Bonnet Gravity}",
    eprint = "1904.09997",
    archivePrefix = "arXiv",
    primaryClass = "gr-qc",
    doi = "10.1103/PhysRevLett.123.011101",
    journal = "Phys. Rev. Lett.",
    volume = "123",
    number = "1",
    pages = "011101",
    year = "2019"
}

@article{Konoplya:2020bxa,
    author = "Konoplya, R. A. and Zinhailo, A. F.",
    title = "{Quasinormal modes, stability and shadows of a black hole in the 4D Einstein\textendash{}Gauss\textendash{}Bonnet gravity}",
    eprint = "2003.01188",
    archivePrefix = "arXiv",
    primaryClass = "gr-qc",
    doi = "10.1140/epjc/s10052-020-08639-8",
    journal = "Eur. Phys. J. C",
    volume = "80",
    number = "11",
    pages = "1049",
    year = "2020"
}

@article{Perlick:2015vta,
    author = "Perlick, Volker and Tsupko, Oleg Yu. and Bisnovatyi-Kogan, Gennady S.",
    title = "{Influence of a plasma on the shadow of a spherically symmetric black hole}",
    eprint = "1507.04217",
    archivePrefix = "arXiv",
    primaryClass = "gr-qc",
    doi = "10.1103/PhysRevD.92.104031",
    journal = "Phys. Rev. D",
    volume = "92",
    number = "10",
    pages = "104031",
    year = "2015"
}

@article{Amarilla:2011fx,
    author = "Amarilla, Leonardo and Eiroa, Ernesto F.",
    title = "{Shadow of a rotating braneworld black hole}",
    eprint = "1112.6349",
    archivePrefix = "arXiv",
    primaryClass = "gr-qc",
    doi = "10.1103/PhysRevD.85.064019",
    journal = "Phys. Rev. D",
    volume = "85",
    pages = "064019",
    year = "2012"
}

@article{Vagnozzi:2019apd,
    author = "Vagnozzi, Sunny and Visinelli, Luca",
    title = "{Hunting for extra dimensions in the shadow of M87*}",
    eprint = "1905.12421",
    archivePrefix = "arXiv",
    primaryClass = "gr-qc",
    doi = "10.1103/PhysRevD.100.024020",
    journal = "Phys. Rev. D",
    volume = "100",
    number = "2",
    pages = "024020",
    year = "2019"
}

@article{Abdujabbarov:2016hnw,
    author = "Abdujabbarov, Ahmadjon and Amir, Muhammed and Ahmedov, Bobomurat and Ghosh, Sushant G.",
    title = "{Shadow of rotating regular black holes}",
    eprint = "1604.03809",
    archivePrefix = "arXiv",
    primaryClass = "gr-qc",
    doi = "10.1103/PhysRevD.93.104004",
    journal = "Phys. Rev. D",
    volume = "93",
    number = "10",
    pages = "104004",
    year = "2016"
}

@article{Grenzebach:2014fha,
    author = {Grenzebach, Arne and Perlick, Volker and L\"ammerzahl, Claus},
    title = "{Photon Regions and Shadows of Kerr-Newman-NUT Black Holes with a Cosmological Constant}",
    eprint = "1403.5234",
    archivePrefix = "arXiv",
    primaryClass = "gr-qc",
    doi = "10.1103/PhysRevD.89.124004",
    journal = "Phys. Rev. D",
    volume = "89",
    number = "12",
    pages = "124004",
    year = "2014"
}

@article{Gralla:2019xty,
    author = "Gralla, Samuel E. and Holz, Daniel E. and Wald, Robert M.",
    title = "{Black Hole Shadows, Photon Rings, and Lensing Rings}",
    eprint = "1906.00873",
    archivePrefix = "arXiv",
    primaryClass = "astro-ph.HE",
    doi = "10.1103/PhysRevD.100.024018",
    journal = "Phys. Rev. D",
    volume = "100",
    number = "2",
    pages = "024018",
    year = "2019"
}

@article{Abramowicz:2011xu,
    author = "Abramowicz, Marek A. and Fragile, P. Chris",
    title = "{Foundations of Black Hole Accretion Disk Theory}",
    eprint = "1104.5499",
    archivePrefix = "arXiv",
    primaryClass = "astro-ph.HE",
    reportNumber = "NSF-KITP-12-055",
    doi = "10.12942/lrr-2013-1",
    journal = "Living Rev. Rel.",
    volume = "16",
    pages = "1",
    year = "2013"
}

@article{1979AA....75..228L,
       author = "Luminet, J. -P.",
        title = "{Image of a spherical black hole with thin accretion disk.}",
      journal ="Astronomy and Astrophysics",
         year = "{1979}",
        month = "may",
       volume = "75",
        pages = "228-235",
       adsurl = "https://ui.adsabs.harvard.edu/abs/1979AA....75..228L"
}

@ARTICLE{EHT_M87_PaperVI,
   author = {{Event Horizon Telescope Collaboration} and {Akiyama}, K. and
	{Alberdi}, A. and {Alef}, W. and {Asada}, K. and {Azulay}, R. and
	{Baczko}, A.-K. and {Ball}, D. and {Balokovi{\'c}}, M. and {Barrett}, J. and others},
    title = "{First M87 Event Horizon Telescope Results. VI. The Shadow and Mass of the Central Black Hole}",
  journal = {Astrophys. J. Lett.},
     year = 2019,
    month = apr,
   volume = 875,
      eid = {L6},
    pages = {L6},
      doi = {10.3847/2041-8213/ab1141},
   adsurl = {http://adsabs.harvard.edu/abs/2019ApJ...875L...6E}
}

@ARTICLE{EHT_SgrA_PaperVI,
	author = {{Event Horizon Telescope Collaboration} and {Akiyama}, Kazunori and others},
        title = "{First Sagittarius A* Event Horizon Telescope Results. VI. Testing the Black Hole Metric}",
      journal = {Astrophys. J. Lett.},
         year = 2022,
        month = may,
       volume = {930},
       number = {2},
          eid = {L17},
        pages = {L17},
          doi = {10.3847/2041-8213/ac6756},
       adsurl = {https://ui.adsabs.harvard.edu/abs/2022ApJ...930L..17E}
}

@article{Narayan:2013gca,
    author = "Narayan, Ramesh and McClintock, Jeffrey E.",
    title = "{Observational Evidence for Black Holes}",
    eprint = "1312.6698",
    archivePrefix = "arXiv",
    primaryClass = "astro-ph.HE",
    month = "12",
    year = "2013"
}

@article{Falcke:1999pj,
    author = "Falcke, Heino and Melia, Fulvio and Agol, Eric",
    title = "{Viewing the shadow of the black hole at the galactic center}",
    eprint = "astro-ph/9912263",
    archivePrefix = "arXiv",
    reportNumber = "HFA-EPRINT-NO-33",
    doi = "10.1086/312423",
    journal = "Astrophys. J. Lett.",
    volume = "528",
    pages = "L13",
    year = "2000"
}

@ARTICLE{Fishbone76,
       author = {{Fishbone}, L.~G. and {Moncrief}, V.},
        title = "{Relativistic fluid disks in orbit around Kerr black holes}",
      journal = {Astrophys. J.},
         year = 1976,
        month = aug,
       volume = {207},
       number = {1},
          eid = {962-976},
        pages = {962-976},
          doi = {10.1086/154565},
       adsurl = {https://ui.adsabs.harvard.edu/abs/1976ApJ...207..962F}
}

@ARTICLE{Font02a,
       author = {{Font}, Jos{\'e} A. and {Daigne}, Fr{\'e}d{\'e}ric},
        title = "{The runaway instability of thick discs around black holes - I. The constant angular momentum case}",
      journal = {Mon. Not. R. Astron. Soc.},
         year = 2002,
        month = aug,
       volume = {334},
       number = {2},
          eid = {383-400},
        pages = {383-400},
          doi = {10.1046/j.1365-8711.2002.05515.x},
archivePrefix = {arXiv},
       eprint = {astro-ph/0203403},
 primaryClass = {astro-ph},
       adsurl = {https://ui.adsabs.harvard.edu/abs/2002MNRAS.334..383F}
}

@article{Moscibrodzka2016,
	Adsurl = {http://adsabs.harvard.edu/abs/2016A%26A...586A..38M},
	Archiveprefix = {arXiv},
	Author = {{Mo{\'s}cibrodzka}, M. and {Falcke}, H. and {Shiokawa}, H.},
	Doi = {10.1051/0004-6361/201526630},
	Eid = {A38},
	Eprint = {1510.07243},
	Journal = {Astron. Astrophys.},
	Month = feb,
	Pages = {A38},
	Primaryclass = {astro-ph.HE},
	Title = {{General relativistic magnetohydrodynamical simulations of the jet in M 87}},
	Volume = 586,
	Year = 2016}

@article{Gimeno_Soler_2017,
   title={Magnetised Polish doughnuts revisited},
   volume={607},
   ISSN={1432-0746},
   url={http://dx.doi.org/10.1051/0004-6361/201730935},
   DOI={10.1051/0004-6361/201730935},
   journal={Astronomy and Astrophysics},
   publisher={EDP Sciences},
   author={Gimeno-Soler, Sergio and Font, José A.},
   year={2017},
   month={Nov},
   pages={A68}
}

@ARTICLE{Gimeno2024,
       author = {{Gimeno-Soler}, Sergio and {Pimentel}, Oscar M. and {Lora-Clavijo}, Fabio D. and {Cruz-Osorio}, Alejandro and {Font}, Jos{\'e} A.},
        title = "{Magnetized tori with magnetic polarization around Kerr black holes: Variable angular momentum disks}",
      journal = {\prd},
         year = 2024,
        month = jul,
       volume = {110},
       number = {2},
          eid = {023023},
        pages = {023023},
          doi = {10.1103/PhysRevD.110.023023},
archivePrefix = {arXiv},
       eprint = {2303.15867},
 primaryClass = {astro-ph.HE},
       adsurl = {https://ui.adsabs.harvard.edu/abs/2024PhRvD.110b3023G}
}

@article{Lahiri:2020sza,
    author = "Lahiri, Sayantani and Gimeno-Soler, Sergio and Font, Jos\'e A. and Mej\'\i{}as, Alejandro Mus",
    title = "{Stationary models of magnetized viscous tori around a Schwarzschild black hole}",
    eprint = "2012.06835",
    archivePrefix = "arXiv",
    primaryClass = "gr-qc",
    doi = "10.1103/PhysRevD.103.044034",
    journal = "Phys. Rev. D",
    volume = "103",
    number = "4",
    pages = "044034",
    year = "2021"
}

@ARTICLE{Liska2018b,
       author = {{Liska}, M. and {Tchekhovskoy}, A. and {Quataert}, E.},
        title = "{Large-scale poloidal magnetic field dynamo leads to powerful jets in GRMHD simulations of black hole accretion with toroidal field}",
      journal = {Monthly Notices of the Royal Astronomical Society},
         year = 2020,
        month = may,
       volume = {494},
       number = {3},
        pages = {3656-3662},
          doi = {10.1093/mnras/staa955},
archivePrefix = {arXiv},
       eprint = {1809.04608},
 primaryClass = {astro-ph.HE},
       adsurl = {https://ui.adsabs.harvard.edu/abs/2020MNRAS.494.3656L}
}

@ARTICLE{Pimentel2021,
       author = {{Pimentel Diaz}, Oscar M. and {Fragile}, P. Chris and {Lora-Clavijo}, F.~D. and {Ierace}, Bridget and {Bollimpalli}, Deepika},
        title = "{Magneto-rotational instability in magnetically polarized discs}",
      journal = {Monthly Notices of the Royal Astronomical Society},
         year = 2021,
        month = aug,
       volume = {505},
       number = {3},
        pages = {4278-4288},
          doi = {10.1093/mnras/stab1520},
archivePrefix = {arXiv},
       eprint = {2105.11329},
 primaryClass = {astro-ph.HE},
       adsurl = {https://ui.adsabs.harvard.edu/abs/2021MNRAS.505.4278P}
}

@BOOK{2002apa..book.....F,
       author = {{Frank}, Juhan and {King}, Andrew and {Raine}, Derek J.},
        title = "{Accretion Power in Astrophysics: Third Edition}",
         year = 2002,
       adsurl = {https://ui.adsabs.harvard.edu/abs/2002apa..book.....F}
}

@ARTICLE{Kozlowski:1978,
   author = {{Kozlowski}, M. and {Jaroszynski}, M. and {Abramowicz}, M.~A.
  },
    title = "{The analytic theory of fluid disks orbiting the Kerr black hole}",
  journal = {Astronomy and Astrophysics},
     year = 1978,
    month = feb,
   volume = 63,
    pages = {209-220},
   adsurl = {http://adsabs.harvard.edu/abs/1978A%26A....63..209K}
}

@ARTICLE{Abramowicz:1978,
   author = {{Abramowicz}, M. and {Jaroszynski}, M. and {Sikora}, M.},
    title = "{Relativistic, accreting disks}",
  journal = {Astronomy and Astrophysics},
     year = 1978,
    month = feb,
   volume = 63,
    pages = {221-224},
   adsurl = {http://adsabs.harvard.edu/abs/1978A%26A....63..221A}
}

@ARTICLE{Olivares2020,
       author = {{Olivares}, Hector and {Younsi}, Ziri and {Fromm}, Christian M. and
         {De Laurentis}, Mariafelicia and {Porth}, Oliver and {Mizuno}, Yosuke and
         {Falcke}, Heino and {Kramer}, Michael and {Rezzolla}, Luciano},
        title = "{How to tell an accreting boson star from a black hole}",
      journal = {MNRAS},
         year = 2020,
        month = jul,
       volume = {497},
       number = {1},
        pages = {521-535},
          doi = {10.1093/mnras/staa1878},
archivePrefix = {arXiv},
       eprint = {1809.08682},
 primaryClass = {gr-qc},
       adsurl = {https://ui.adsabs.harvard.edu/abs/2020MNRAS.497..521O}
}

@ARTICLE{Komissarov:2006,
       author = {{Komissarov}, S.~S.},
        title = "{Magnetized tori around Kerr black holes: analytic solutions with a toroidal magnetic field}",
      journal = {Monthly Notices of the Royal Astronomical Society},
         year = 2006,
        month = may,
       volume = {368},
       number = {3},
        pages = {993-1000},
          doi = {10.1111/j.1365-2966.2006.10183.x},
archivePrefix = {arXiv},
       eprint = {astro-ph/0601678},
 primaryClass = {astro-ph},
       adsurl = {https://ui.adsabs.harvard.edu/abs/2006MNRAS.368..993K}
}

@article{Lahiri_2020,
   title={Second order causal hydrodynamics in Eckart frame: using gradient expansion scheme},
   volume={37},
   ISSN={1361-6382},
   url={http://dx.doi.org/10.1088/1361-6382/ab712f},
   DOI={10.1088/1361-6382/ab712f},
   number={7},
   journal={Classical and Quantum Gravity},
   publisher={IOP Publishing},
   author={Lahiri, Sayantani},
   year={2020},
   month={Mar},
   pages={075010}
}

@article{Muller:1967zza,
    author = "Muller, Ingo",
    title = "{Zum Paradoxon der Warmeleitungstheorie}",
    doi = "10.1007/BF01326412",
    journal = "Z. Phys.",
    volume = "198",
    pages = "329--344",
    year = "1967"
}

@article{Israel:1976tn,
    author = "Israel, W.",
    title = "{Nonstationary irreversible thermodynamics: A Causal relativistic theory}",
    doi = "10.1016/0003-4916(76)90064-6",
    journal = "Annals Phys.",
    volume = "100",
    pages = "310--331",
    year = "1976"
}

@article{Johannsen2010,
	Adsurl = {http://adsabs.harvard.edu/abs/2010ApJ...718..446J},
	Archiveprefix = {arXiv},
	Author = {{Johannsen}, T. and {Psaltis}, D.},
	Doi = {10.1088/0004-637X/718/1/446},
	Eprint = {1005.1931},
	Journal = {Astrophys. J.},
	Month = jul,
	Pages = {446-454},
	Primaryclass = {astro-ph.HE},
	Title = {{Testing the No-hair Theorem with Observations in the Electromagnetic Spectrum. II. Black Hole Images}},
	Volume = 718,
	Year = 2010}

@article{ISRAEL1981204,
title = "Thermodynamics of relativistic systems",
journal = "Physica A: Statistical Mechanics and its Applications",
volume = "106",
number = "1",
pages = "204 - 214",
year = "1981",
issn = "0378-4371",
doi = "https://doi.org/10.1016/0378-4371(81)90220-X",
url = "http://www.sciencedirect.com/science/article/pii/037843718190220X"
}

@book{anile2005relativistic,
  title={Relativistic fluids and magneto-fluids: With applications in astrophysics and plasma physics},
  author={Anile, Angelo Marcello},
  year={2005},
  publisher={Cambridge University Press}
}

@ARTICLE{Fromm2021b,
       author = {{Fromm}, Christian M. and {Cruz-Osorio}, Alejandro and {Mizuno}, Yosuke and {Nathanail}, Antonios and {Younsi}, Ziri and {Porth}, Oliver and {Olivares}, Hector and {Davelaar}, Jordy and {Falcke}, Heino and {Kramer}, Michael and {Rezzolla}, Luciano},
        title = "{Impact of non-thermal particles on the spectral and structural properties of M87}",
      journal = {Astron. Astrophys.},
         year = 2022,
        month = apr,
       volume = {660},
          eid = {A107},
        pages = {A107},
          doi = {10.1051/0004-6361/202142295},
archivePrefix = {arXiv},
       eprint = {2111.02518},
 primaryClass = {astro-ph.HE},
       adsurl = {https://ui.adsabs.harvard.edu/abs/2022A&A...660A.107F}}

@ARTICLE{EHT_M87_PaperI,
   author = {{Event Horizon Telescope Collaboration} and {Akiyama}, K. and others},
    title = "{First M87 Event Horizon Telescope Results. I. The Shadow of the Supermassive Black Hole}",
  journal = {Astrophys. J. Lett.},
     year = 2019,
    month = apr,
   volume = 875,
      eid = {L1},
    pages = {L1},
      doi = {10.3847/2041-8213/ab0ec7},
   adsurl = {http://adsabs.harvard.edu/abs/2019ApJ...875L...1E}
}

@ARTICLE{EHT_M87_PaperV,
   author = {{Event Horizon Telescope Collaboration} and {Akiyama}, K. and others},
    title = "{First M87 Event Horizon Telescope Results. V. Physical Origin of the Asymmetric Ring}",
  journal = {Astrophys. J. Lett.},
     year = 2019,
    month = apr,
   volume = 875,
      eid = {L5},
    pages = {L5},
      doi = {10.3847/2041-8213/ab0f43},
   adsurl = {http://adsabs.harvard.edu/abs/2019ApJ...875L...5E}
}

@ARTICLE{EHT_M87_2018,
  author = {{Event Horizon Telescope Collaboration} and {Akiyama}, K. and others},
        title = "{The persistent shadow of the supermassive black hole of M87: II. Model comparisons and theoretical interpretations}",
	journal = {Astron. Astrophys.},
	year = 2025,
	month = jan,
	volume = {693},
	eid = {A265},
	pages = {A265},
	doi = {10.1051/0004-6361/202451296},
	adsurl = {https://ui.adsabs.harvard.edu/abs/2025A&A...693A.265E}
	}

@ARTICLE{EHT_SgrA_PaperI,
       author = {{Event Horizon Telescope Collaboration} and {Akiyama},
                  K. and others},
        title = "{First Sagittarius A* Event Horizon Telescope Results. I. The Shadow of the Supermassive Black Hole in the Center of the Milky Way}",
      journal = {Astrophys. J. Lett.},
         year = 2022,
        month = may,
       volume = {930},
       number = {2},
          eid = {L12},
        pages = {L12},
          doi = {10.3847/2041-8213/ac6674},
       adsurl = {https://ui.adsabs.harvard.edu/abs/2022ApJ...930L..12A}}

@ARTICLE{EHT_SgrA_PaperV_etal,
        author = {{Event Horizon Telescope Collaboration} and {Akiyama}, K. and others},
        title = "{First Sagittarius A* Event Horizon Telescope Results. V. Testing Astrophysical Models of the Galactic Center Black Hole}",
      journal = {Astrophys. J. Lett.},
         year = 2022,
        month = may,
       volume = {930},
       number = {2},
          eid = {L16},
        pages = {L16},
          doi = {10.3847/2041-8213/ac6672},
       adsurl = {https://ui.adsabs.harvard.edu/abs/2022ApJ...930L..16A}}

@ARTICLE{Wang2025,
       author = {{Wang}, Xinyu and {Zhao}, Zhixing and {Zeng}, Xiao-Xiong and {Wang}, Xin-Yang},
        title = "{Revisiting the shadow of Johannsen-Psaltis black holes}",
      journal = {\prd},
         year = 2025,
        month = apr,
       volume = {111},
       number = {8},
          eid = {084054},
        pages = {084054},
          doi = {10.1103/PhysRevD.111.084054},
archivePrefix = {arXiv},
       eprint = {2501.08287},
 primaryClass = {gr-qc},
       adsurl = {https://ui.adsabs.harvard.edu/abs/2025PhRvD.111h4054W}
}

@ARTICLE{Younsi2016,
       author = {{Younsi}, Ziri and {Zhidenko}, Alexander and {Rezzolla}, Luciano and {Konoplya}, Roman and {Mizuno}, Yosuke},
        title = "{New method for shadow calculations: Application to parametrized axisymmetric black holes}",
      journal = {\prd},
         year = 2016,
        month = oct,
       volume = {94},
       number = {8},
          eid = {084025},
        pages = {084025},
          doi = {10.1103/PhysRevD.94.084025},
archivePrefix = {arXiv},
       eprint = {1607.05767},
 primaryClass = {gr-qc},
       adsurl = {https://ui.adsabs.harvard.edu/abs/2016PhRvD..94h4025Y}
}

@ARTICLE{Younsi2023,
       author = {{Younsi}, Ziri and {Psaltis}, Dimitrios and {{\"O}zel}, Feryal},
        title = "{Black Hole Images as Tests of General Relativity: Effects of Spacetime Geometry}",
      journal = {\apj},
         year = 2023,
        month = jan,
       volume = {942},
       number = {1},
          eid = {47},
        pages = {47},
          doi = {10.3847/1538-4357/aca58a},
archivePrefix = {arXiv},
       eprint = {2111.01752},
 primaryClass = {astro-ph.HE},
       adsurl = {https://ui.adsabs.harvard.edu/abs/2023ApJ...942...47Y}
}

@ARTICLE{Chatterjee2025,
       author = {{Chatterjee}, Koushik and {Younsi}, Ziri and {Kocherlakota}, Prashant and {Narayan}, Ramesh},
        title = "{On the Universality of Energy Extraction from Black Hole Spacetimes}",
      journal = {\apjl},
         year = 2025,
        month = oct,
       volume = {991},
       number = {2},
          eid = {L58},
        pages = {L58},
          doi = {10.3847/2041-8213/ae0740},
archivePrefix = {arXiv},
       eprint = {2310.20043},
 primaryClass = {gr-qc},
       adsurl = {https://ui.adsabs.harvard.edu/abs/2025ApJ...991L..58C}
}

@ARTICLE{Chatterjee2023,
       author = {{Chatterjee}, Koushik and {Kocherlakota}, Prashant and {Younsi}, Ziri and {Narayan}, Ramesh},
        title = "{Energy Extraction from Spinning Stringy Black Holes}",
      journal = {arXiv e-prints},
         year = 2023,
        month = oct,
          eid = {arXiv:2310.20040},
        pages = {arXiv:2310.20040},
          doi = {10.48550/arXiv.2310.20040},
archivePrefix = {arXiv},
       eprint = {2310.20040},
 primaryClass = {gr-qc},
       adsurl = {https://ui.adsabs.harvard.edu/abs/2023arXiv231020040C}
}


   
    \end{document}